\documentclass[12pt]{iopart}

\usepackage{iopams}
\usepackage{graphicx}
\usepackage{epstopdf}
\usepackage{siunitx}
\usepackage{subcaption}

\begin{document}
\title[Resonant fast-ion redistribution in MAST]{Measurements and modelling of fast-ion redistribution due to resonant MHD instabilities in MAST}
\author{O M Jones$^{1,2}$, M Cecconello$^3$, K G McClements$^2$, I Klimek$^3$, R J Akers$^2$, W U Boeglin$^4$, D L Keeling$^2$, A J Meakins$^2$, R V Perez$^4$, S E Sharapov$^2$, M Turnyanskiy$^5$, and the MAST team$^2$}
\address{$^1$Department of Physics, Durham University, South Road, Durham DH1 3LE, UK}
\address{$^2$CCFE, Culham Science Centre, Abingdon, Oxon. OX14 3DB, UK}
\address{$^3$Department of Physics and Astronomy, Uppsala University, SE-751 05 Uppsala, Sweden}
\address{$^4$Department of Physics, Florida International University, 11200 SW 8 ST, CP204, Miami, Florida 33199, USA}
\address{$^5$ITER Physics Department, EUROfusion PMU Garching, Boltzmannstra{\ss}e 2, D-85748 Garching, Germany}

\ead{owen.jones@ccfe.ac.uk}

\begin{abstract}
The results of a comprehensive investigation into the effects of toroidicity-induced Alfv\'{e}n eigenmodes (TAE) and energetic particle modes on the NBI-generated fast-ion population in MAST plasmas are reported. Fast-ion redistribution due to frequency-chirping TAE in the range \SIrange{50}{100}{kHz}, and frequency-chirping energetic particle modes known as fishbones in the range \SIrange{20}{50}{kHz}, is observed. TAE and fishbones are also observed to cause losses of fast ions from the plasma. The spatial and temporal evolution of the fast-ion distribution is determined using a fission chamber, a radially-scanning collimated neutron flux monitor, a fast-ion deuterium alpha spectrometer and a charged fusion product detector. Modelling using the global transport analysis code \textsc{Transp}, with \emph{ad hoc} anomalous diffusion and fishbone loss models introduced, reproduces the coarsest features of the affected fast-ion distribution in the presence of energetic-particle-driven modes. The spectrally and spatially resolved measurements show however that these models do not fully capture the effects of chirping modes on the fast-ion distribution.
\end{abstract}
\clearpage

\section{Introduction}
\label{sec:Intro}
The Mega-Amp Spherical Tokamak (MAST) was reliant on neutral beam heating to access high-performance operating regimes \cite{Akers2003, Buttery2004}. One of the world's largest spherical tokamaks to date, typical parameters of MAST were: major and minor radius $R=\SI{0.95}{m}$, $a=\SI{0.60}{m}$; plasma current $I_\mathrm{p}=400-\SI{900}{kA}$; toroidal field on axis $B_\mathrm{T}=0.40-\SI{0.58}{T}$; core electron density and temperature $n_{\mathrm{e}0}=\SI{3e19}{m^{-3}}$ and $T_{\mathrm{e}0}=\SI{1}{keV}$. The neutral beam injection (NBI) system installed on MAST consisted of two positive ion sources capable of accelerating deuterons to energies of \SI{75}{keV}. Each injector could deliver up to \SI{2.5}{MW} of power when tuned to optimum perveance \cite{Gee2005}.

Fast ions (FI) in MAST were produced by ionisation of beam neutrals as they propagated through the plasma. The relatively weak magnetic field resulted in large Larmor radii of up to \SI{10}{cm} for the highest energy beam ions. Once they are deposited, FI interact with the background plasma, slowing down by electron drag and scattering off thermal ions. Ultimately, the ions may either: thermalise, reaching equilibrium with the background plasma; undergo fusion reactions, producing a triton and proton, or $^3$He nucleus and neutron; become re-neutralised by charge exchange with beam or thermal neutrals; collisionally scatter onto unconfined orbits; or be lost from the plasma due to instabilities or static field perturbations. In order to maximise the electron and ion temperature and fraction of NBI-driven current in the plasma, and hence measures of performance such as plasma beta, it is desirable to confine the FI for as long as possible. Maximising the fraction of NBI-driven current requires these ions to be confined at low plasma density, posing challenges for avoiding energetic particle-driven instabilities. Diagnosing the behaviour of FI in MAST and establishing the mechanisms for their redistribution within and loss from the plasma are the objectives of the present study.

This paper focuses specifically on transport of FI in MAST caused by MHD instabilities which resonate with them \cite{Mikhailovskii1975}. This resonant interaction allows significant energy transfer from the FI population to the MHD mode in the linear growth phase. Subsequent nonlinear evolution of the mode is typically associated with large-scale coherent transport of the resonant FI \cite{Breizman1993}. The mode acts to flatten the FI distribution close to the resonance in real space and in velocity space. In the worst-case scenario, where several resonances overlap or where sufficient free energy is present in the FI distribution to drive the mode over a spatially-extended region, FI may even be transported to the plasma boundary and lost. Such losses, as well as degrading plasma performance, may constitute unacceptable localised heat loads on the plasma facing components in a fusion reactor \cite{Pinches2004}.

MAST provided a useful facility for the study of energetic-particle physics for a number of reasons. Foremost among these were the super-Alfv\'enic beam ions, which allowed Alfv\'en eigenmodes to be driven via the fundamental resonance $v_\parallel=v_A$, and the large orbit widths of the FI which pose a challenge to the verification of numerical models \cite{Pinches2012}. MAST was therefore equipped with a number of advanced diagnostics designed to probe the dynamics of the beam-ion population. Recent work by many of the present authors \cite{Cecconello2015} has highlighted the complementary nature of the different FI diagnostics installed on MAST, presenting an overview of the observed FI transport in the presence of chirping toroidicity-induced Alfv\'en eigenmodes (TAE) and fishbones, sawteeth and the long-lived internal kink mode. This paper forms a natural continuation of that study, and aims to provide a quantitative verification of the key results presented in the prior work. The main findings are the confirmation that anomalous transport and losses of FI accompany both chirping TAE and fishbones; the demonstration that multiple diagnostic signals may in some cases be modelled consistently using anomalous FI diffusion in a global transport model; and the identification of the passing FI population in the core of the plasma as being susceptible to redistribution in the presence of fishbones. This work expands upon the previous study by providing: 
\begin{itemize}
\item{a rigorous quantitative assessment of the degree to which the reduction in FI confinement is correlated with TAE and fishbones;}
\item{direct experimental confirmation of the fact that these MHD modes cause losses of FI from the plasma;}
\item{an improved implementation of the fishbone loss model in \textsc{Transp}, including fast-ion deuterium alpha (FIDA) data as a constraint for the first time.}
\end{itemize}

Section \ref{sec:FIDs} of this paper describes the FI diagnostics used in this study. The nature of the energetic-particle-driven MHD activity commonly observed in MAST is illustrated in section \ref{sec:MHD}. Section \ref{sec:Scenarios} outlines the plasma scenarios investigated in the course of this study, and sections \ref{sec:Results} and \ref{sec:Transp} present the experimental results and associated FI transport modelling. This modelling was performed with the 1.5-dimensional (1D profiles mapped onto a 2D equilibrium grid), time-dependent transport analysis code \textsc{Transp} \cite{Transp}. The Monte Carlo particle-tracking module \textsc{Nubeam} \cite{Pankin2004} was used to model the deposition and orbits of FI and to generate local FI distributions as a function of energy and pitch. The main conclusions of the work are presented in section \ref{sec:Summary}.

\section{Fast-ion diagnostics on MAST}
\label{sec:FIDs}
\subsection{Description of the diagnostics}
\begin{figure}[t]
\centering
\includegraphics[width=\textwidth]{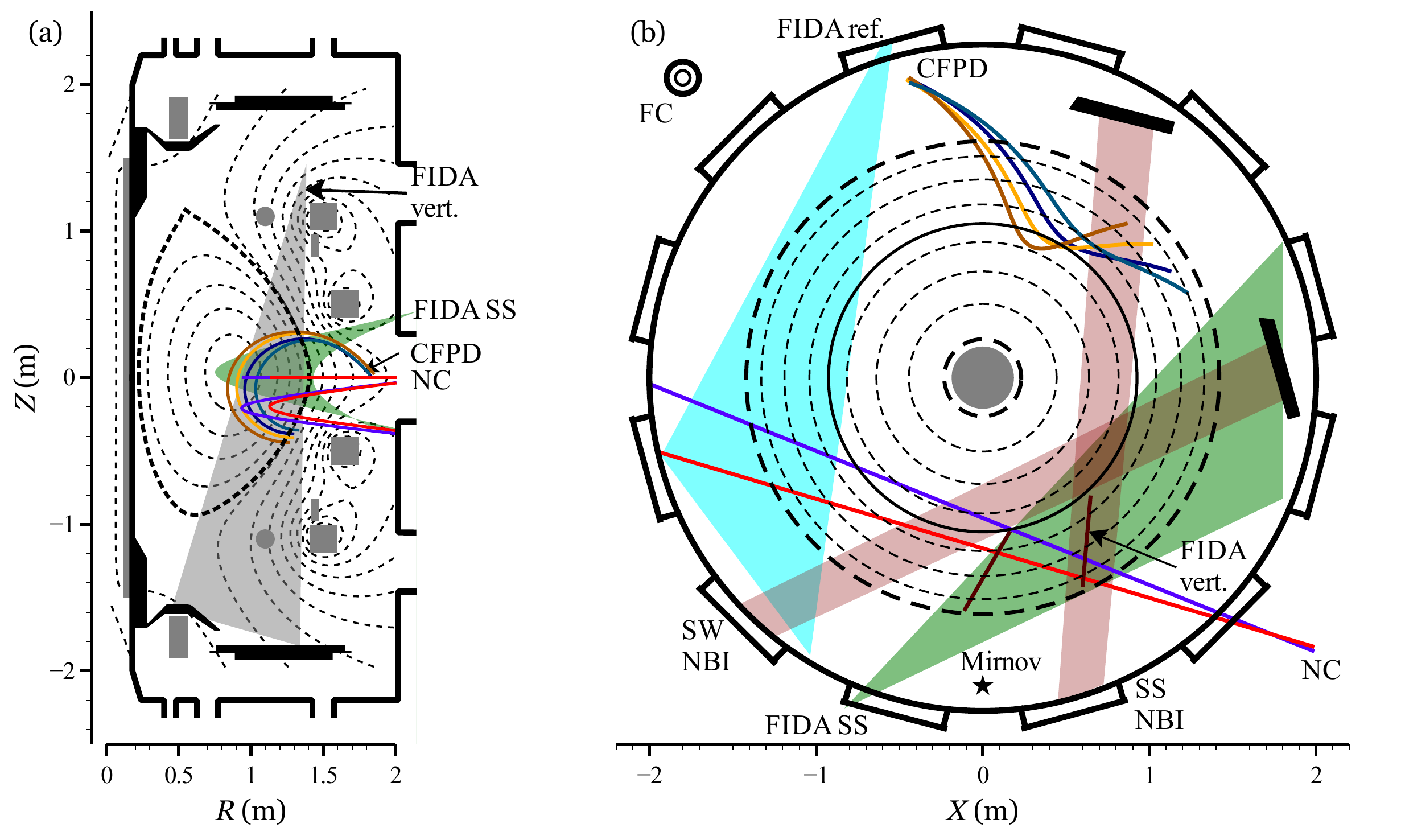}
\caption{Poloidal (a) and equatorial (b) projections of the viewing geometry of each of the MAST FI diagnostics. Flux surfaces are superimposed as dashed lines; solid grey and black components respectively are cross-sections of poloidal field coils and graphite armour on the divertor and beam dumps. Green and grey shaded areas show the regions spanned by toroidally- and vertically-oriented SS beam-viewing FIDA lines of sight (LOS) respectively. Red and blue lines show examples of NC LOS, both midplane and diagonally-oriented chords. Coloured, curved lines show fusion proton/triton trajectories from the plasma which reach the CFPD. NBI footprints are shown as red shaded regions. The cyan shaded area marks the region spanned by toroidal FIDA reference views, used for background subtraction. Also shown in (b) are the positions of the fission chamber (FC) and an outboard midplane Mirnov pick-up coil, used to detect MHD activity in the plasma. Midplane intersection radii of the active (SS beam-viewing) and reference (background subtraction) vertically-oriented FIDA chords are shown as dark brown lines in (b). Adapted from \href{http://iopscience.iop.org/0741-3335/57/1/014006/article}{doi:10.1088/0741-3335/57/1/014006}. \copyright IOP Publishing. Reproduced by permission of IOP Publishing. All rights reserved.}
\label{fig:FIDLOS}
\end{figure}

During the 2013 MAST experimental campaign, four FI diagnostics were available on the device; one of these, the \emph{charged fusion product detector} (CFPD), is a novel prototype diagnostic which was only installed for a limited period \cite{Perez2014}. The other three diagnostics are: a uranium-235 fission chamber for volume-integrated neutron flux measurements \cite{Stammers2006}; a collimated, radially-scanning neutron detector (known as the neutron camera, or NC) \cite{Cecconello2014a}; and a dual-view Fast-Ion Deuterium Alpha (FIDA) diagnostic \cite{Michael2013b}. Figure \ref{fig:FIDLOS} shows the detector positions and lines of sight of each of these diagnostics projected onto the poloidal and equatorial planes of the MAST vessel.

It should be noted that although measurements of the neutron flux are not fundamentally restricted to FI diagnostic applications, the relatively low ion temperature in MAST plasmas (typically $T_\mathrm{i}\lesssim \SI{2}{keV}$ in the core of the plasma) resulted in a very low contribution of thermonuclear fusion reactions to the total neutron yield. Approximately 98\% of D-D fusion reactions in MAST occurred either between a beam ion and a thermal ion (${\sim}80\%$), or between two beam ions (${\sim}20\%$). As a result, the flux of $\SI{2.45}{MeV}$ neutrons leaving the plasma provided information about the non-thermal, high-energy part of the deuterium ion velocity distribution. The $^{235}$U fission chamber, or FC \cite{Stammers2006}, measured the volume-integrated neutron flux from the plasma with a temporal resolution of \SI{10}{\micro\second}, while the NC had four collimated lines of sight (LOS) allowing the detectors to view the plasma through a thin flange in the MAST vacuum vessel \cite{Cecconello2014a}. Two of the LOS lay in the machine midplane, and the other two were oriented diagonally downward such that at the point of their tangency to the flux surfaces they lay approximately $\SI{20}{cm}$ below the midplane. The NC assembly was mounted on a curved rail, allowing the LOS to be scanned across the plasma between discharges; repeated discharges allowed a neutron emission profile, integrated within the NC fields of view, to be obtained as a function of viewing tangency radius. Data are typically integrated over $\SI{1}{ms}$, setting the temporal resolution of the measurements.

The CFPD was installed on MAST to provide a complementary measurement to that provided by the NC \cite{Perez2014}. Four detectors measured the flux of $\SI{3.02}{MeV}$ protons and $\SI{1.01}{MeV}$ tritons (the gyro-radii of these two fusion products being very nearly equal) produced by D-D fusion reactions. Collimators in the protective casing allowed fusion products to reach the detectors, and an aluminium foil prevented detection of soft X-rays. The diagnostic was mounted on the end of a linear reciprocating probe inserted into the MAST vessel at the midplane. This diagnostic technique is practical in small, low-field devices such as MAST because the large gyro-radii of the fusion products render them essentially unconfined by the magnetic field; they leave the plasma within a single gyro-orbit. With a suitable equilibrium reconstruction, orbits of these particles may be tracked `backward in time' from the detector to determine the possible location of their emission. Retraction or insertion of the probe between repeated discharges allowed a spatial scan of the proton/triton emission profile. Protons and tritons are discriminated in post-processing. As in the case of the NC, the integration time of the data is typically set to $\SI{1}{ms}$.

The FIDA diagnostic provided a measurement of the FI density by observing the Balmer-alpha (D$_\alpha$) emission of reneutralised fast deuterons which had undergone charge exchange (CX) with beam or halo neutrals \cite{Michael2013b}. \emph{Halo neutrals} are thermal neutrals created by CX between beam neutrals and thermal ions. The light was coupled via optical fibres to a spectrometer, and spectral resolution was provided by dispersion of a bandpass-filtered portion of this light. Spatial resolution was obtained by using individual fibres for each LOS through the plasma. The spectrum of FIDA light provides information about the distribution of the LOS velocity of the FI at the point at which CX occurred; D$_\alpha$ light from reneutralised ions moving away from the lens is redshifted, while that from ions moving toward the lens is blueshifted. Eleven channels could be connected to the spectrometer at any given time, allowing a radial profile of the FIDA emission to be acquired within a single discharge. Data were acquired with $\SI{0.3}{ms}$ temporal resolution. Two sets of views were available: a near-toroidal set from a lens mounted just above the vessel midplane, and a near-vertical set from a lens mounted inside the vessel, looking vertically downward at the neutral beam. The toroidal views were predominantly sensitive to passing FI, with a large pitch $p=v_\parallel/v$, while the vertical views were more sensitive to trapped FI with smaller values of pitch.

\begin{figure}[h]
\centering
\includegraphics[width=0.75\textwidth]{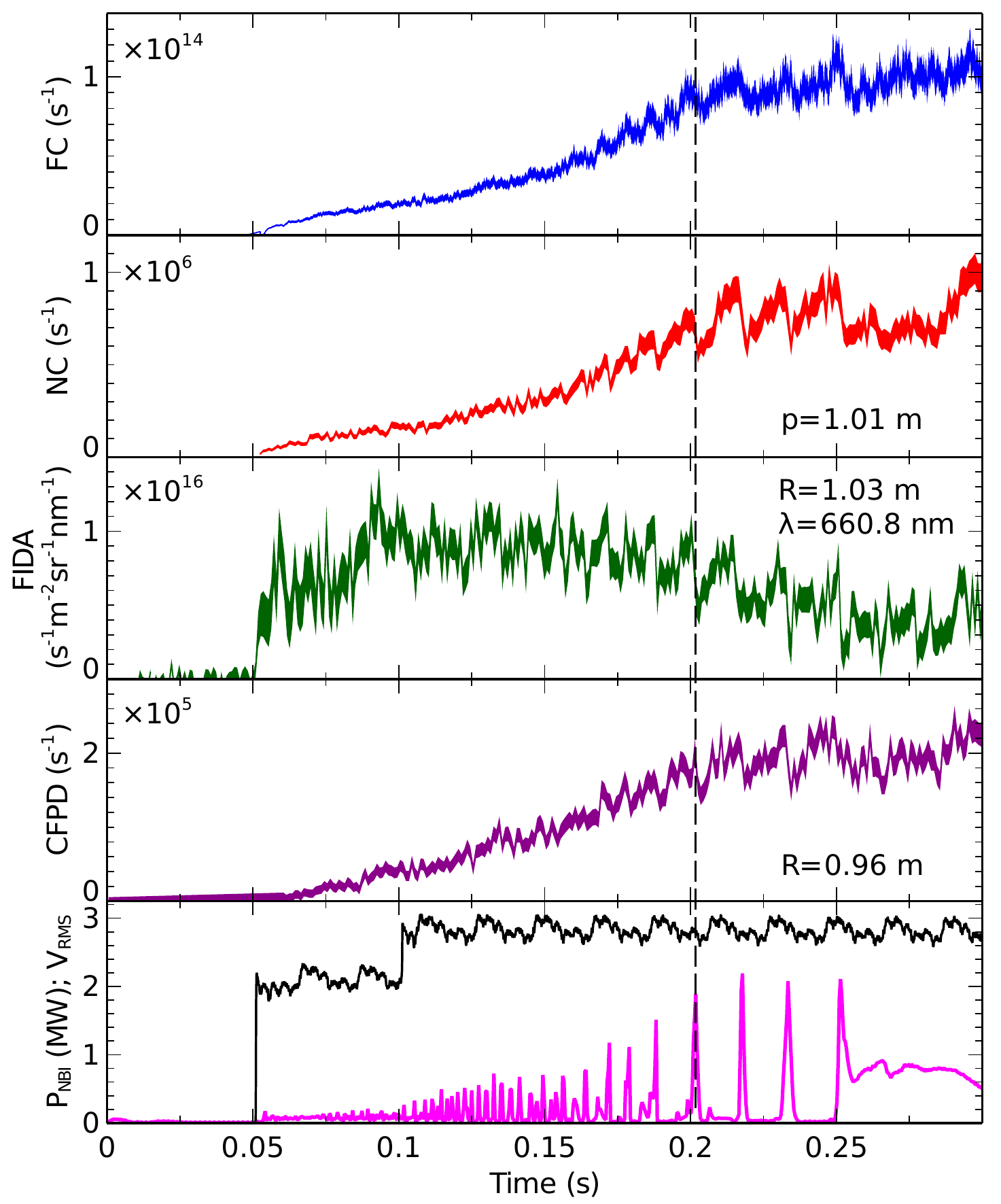}
\caption[Examples of FI diagnostic time traces]{Time traces from single channels of the four MAST FI diagnostics, from shot \#29980. The first four panels show, from top to bottom: fission chamber count rate; neutron camera count rate from a midplane chord with tangency radius $p$; FIDA spectral radiance at wavelength $\lambda$ from a chord intersecting the axis of the SS neutral beam at major radius $R$; CFPD proton count rate from a channel with fusion proton trajectories intersecting the midplane at major radius $R$ (where the values of $p$, $R$ and $\lambda$ are given by the labels in each panel). The thickness of the lines shows the statistical uncertainty in the data. Shown in the bottom panel is the total NBI power and the RMS voltage in vertically-oriented Mirnov coil on the outboard midplane. The dashed vertical line marks the time at which a large fishbone occurred, with a corresponding reduction seen in each of the FI diagnostic signals.}
\label{fig:FIDtraces}
\end{figure}

An example of the data from each of the diagnostics described in this section is shown in figure \ref{fig:FIDtraces}. $\SI{2.0}{MW}$ of neutral beam heating power (at $E_\mathrm{b1}=\SI{70}{keV}$) was applied from $\SI{0.051}{s}$, with an additional $\SI{0.7}{MW}$ (at $E_\mathrm{b2}=\SI{44}{keV}$) applied from $\SI{0.101}{s}$. A series of large fishbones occurred between $\SI{0.20}{s} - \SI{0.25}{s}$; the time of the first of these bursts of MHD activity is indicated by the dashed vertical line. The effect of the fishbones is apparent as a significant drop in the magnitude of the signal in each time trace; this effect is particularly marked in the NC, FIDA and CFPD data, where drops in excess of $25\%$ of the pre-fishbone signal are observed.

\subsection{Velocity-space sensitivity}
\begin{figure}[h]
\centering
\includegraphics[width=0.5\textwidth]{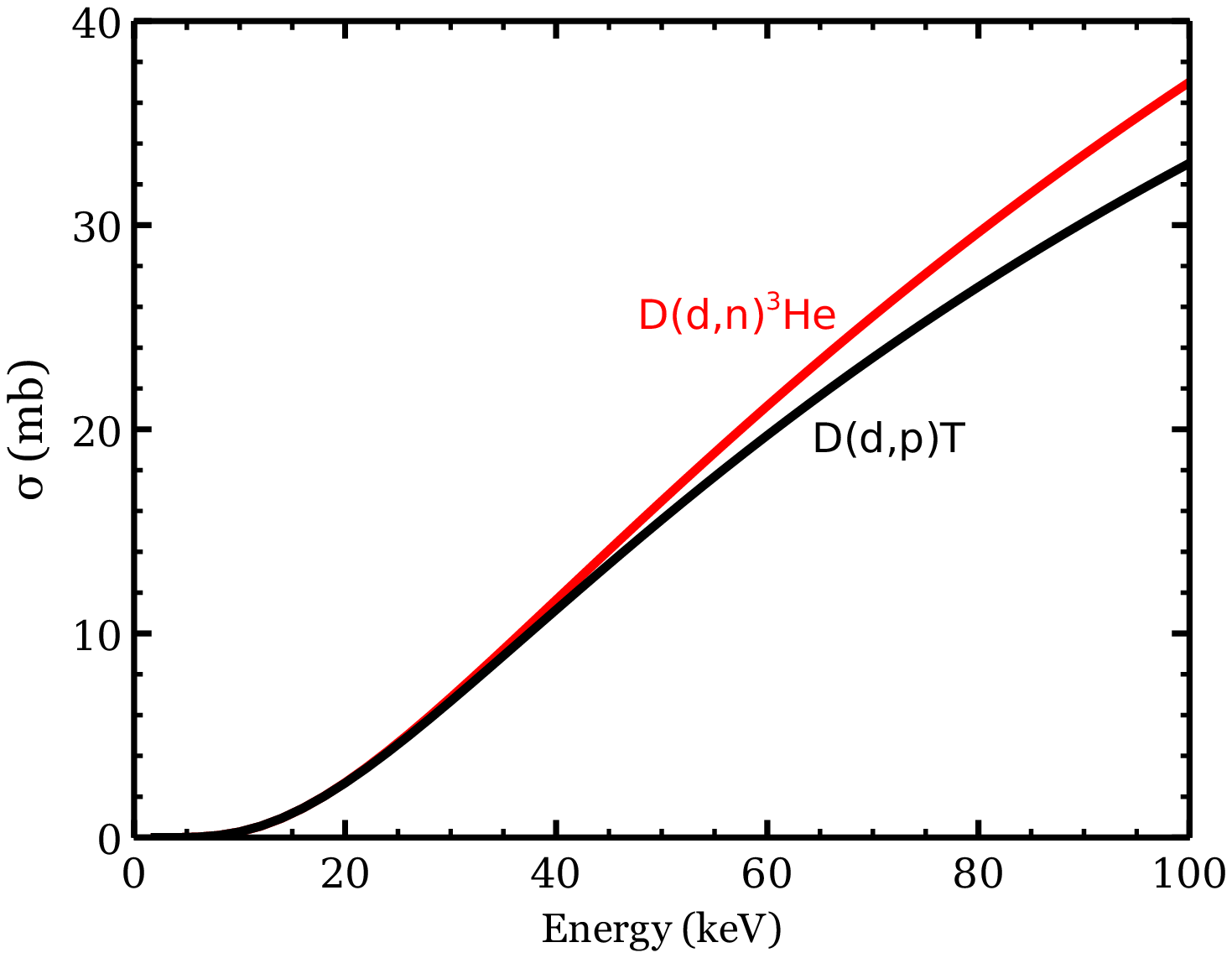}
\caption{Cross-section of the two branches of the D-D fusion reaction as a function of the centre-of-mass energy of the reactants. The empirical formula of Bosch \& Hale (1992) is used to obtain each of the curves \cite{Bosch1992}.}
\label{fig:DDXsec}
\end{figure}

Each of the diagnostics has a sensitivity to FI which varies throughout velocity space as well as real space. This is most apparent in the case of the fusion product diagnostics, since the D-D fusion cross-section is a strong function of reactant energy. Figure \ref{fig:DDXsec} shows the cross-section of the two branches of the D-D fusion reaction as a function of centre-of-mass energy from $\SI{5}{keV}$ to $\SI{100}{keV}$ \cite{Bosch1992}. It is clear that the sensitivity of fusion product measurements is strongly weighted to the high-energy FI, especially given that the contribution of reactions at each energy to the total reactivity is proportional to $\sigma v$ rather than simply to $\sigma$. The weighting of the measurements at each point in velocity space is determined by averaging the reactivity over the velocity distribution of the target ions. This is a thermal Maxwellian distribution in the case of beam-thermal reactions, and a numerically-calculated velocity distribution of beam ions in the case of beam-beam reactions.

\begin{figure}[h]
\centering
\includegraphics[width=0.9\textwidth]{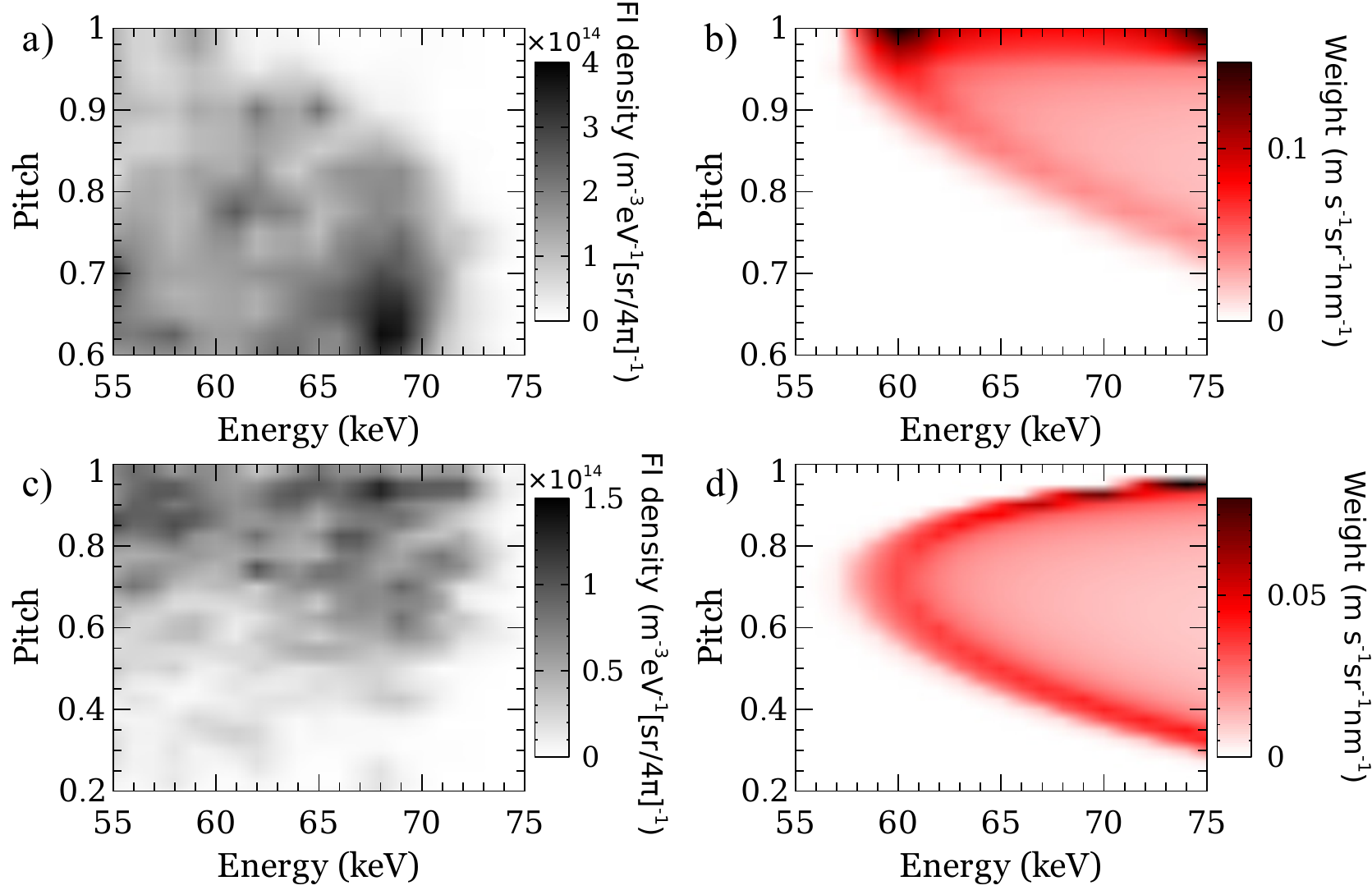}
\caption{(a) High-energy part of the FI distribution from shot \#29976 at $t=\SI{0.2}{s}$, $R=\SI{1.00}{m}$ on the outboard midplane, modelled by \textsc{Transp}. (b) FIDA velocity-space sensitivity function (`weight function') at $\lambda=\SI{661.3}{nm}$ for a toroidally-viewing chord intersecting the neutral beam at $R=\SI{1.00}{m}$. The spectral radiance is obtained by multiplying the FI distribution (panel (a)) by the weight function (panel (b)) and integrating over velocity space. (c) and (d) are similar to (a) and (b), but for a chord intersecting the neutral beam at $R=\SI{1.25}{m}$. Note the different ranges of pitch in (c) and (d) than in (a) and (b).}
\label{fig:wfunc1}
\end{figure}

An illustrative example of the velocity-space sensitivity of the FIDA diagnostic is shown in figure \ref{fig:wfunc1}. FIDA measurements are sensitive only to FI in a bounded region of velocity space; this region is shown in panels (b) and (d) of figure \ref{fig:wfunc1} for two toroidally-viewing chords with different beam intersection radii. Variation of the sensitivity within this bounded region arises due to geometric factors, which determine the proportion of the gyro-orbit in which the LOS velocity of each FI renders it detectable in a particular viewing chord at a particular wavelength. Atomic physics factors also play a part, by determining the probability that interaction with beam or halo neutrals will give rise to the emission of a D$_\alpha$ photon. This geometric part of this sensitivity function, known as a \emph{weight function}, is determined by the geometry of the LOS, the magnetic field and the neutral beam. The CX probability depends on the electron temperature and density via their effect on the beam and halo neutral density and distribution of excited states. NBI energy and full, half and third-energy current fractions also affect the CX probability via their effect on the relative velocity between FI and beam neutrals. Observing FIDA light at a given wavelength corresponds to observing emission which is a product of the wavelength-dependent weight function and the FI distribution at the location of beam intersection. Stark splitting of the Balmer-alpha radiation also plays a part in determining the precise details of the weight function. Given the complexity of their construction, these weight functions must be calculated numerically in a similar manner to the fusion neutron or proton emissivity profiles. The general method for quantitative calculation of FIDA weight functions is detailed by Salewski \emph{et al.} in ref. \cite{Salewski2014}, while the \textsc{FIDAsim} code calculates weight functions based on the contribution of Monte Carlo marker particles, representing reneutralised FI, to the FIDA spectra \cite{Geiger2012}.

With the velocity-space sensitivity of each of the FI diagnostics being determined, diagnostic signals may be modelled based on synthetic FI distributions. These `synthetic diagnostics' rely on measured profiles of plasma density and temperature, rotation velocity and impurity density. Such modelling forms the basis of section \ref{sec:Transp}. Before presenting the experimental results however, we turn our attention to the classification of energetic-particle-driven MHD instabilities typically observed in MAST discharges.

\section{MHD activity observed in MAST}
\label{sec:MHD}
\begin{figure}[h]
\centering
\includegraphics[width=0.95\textwidth]{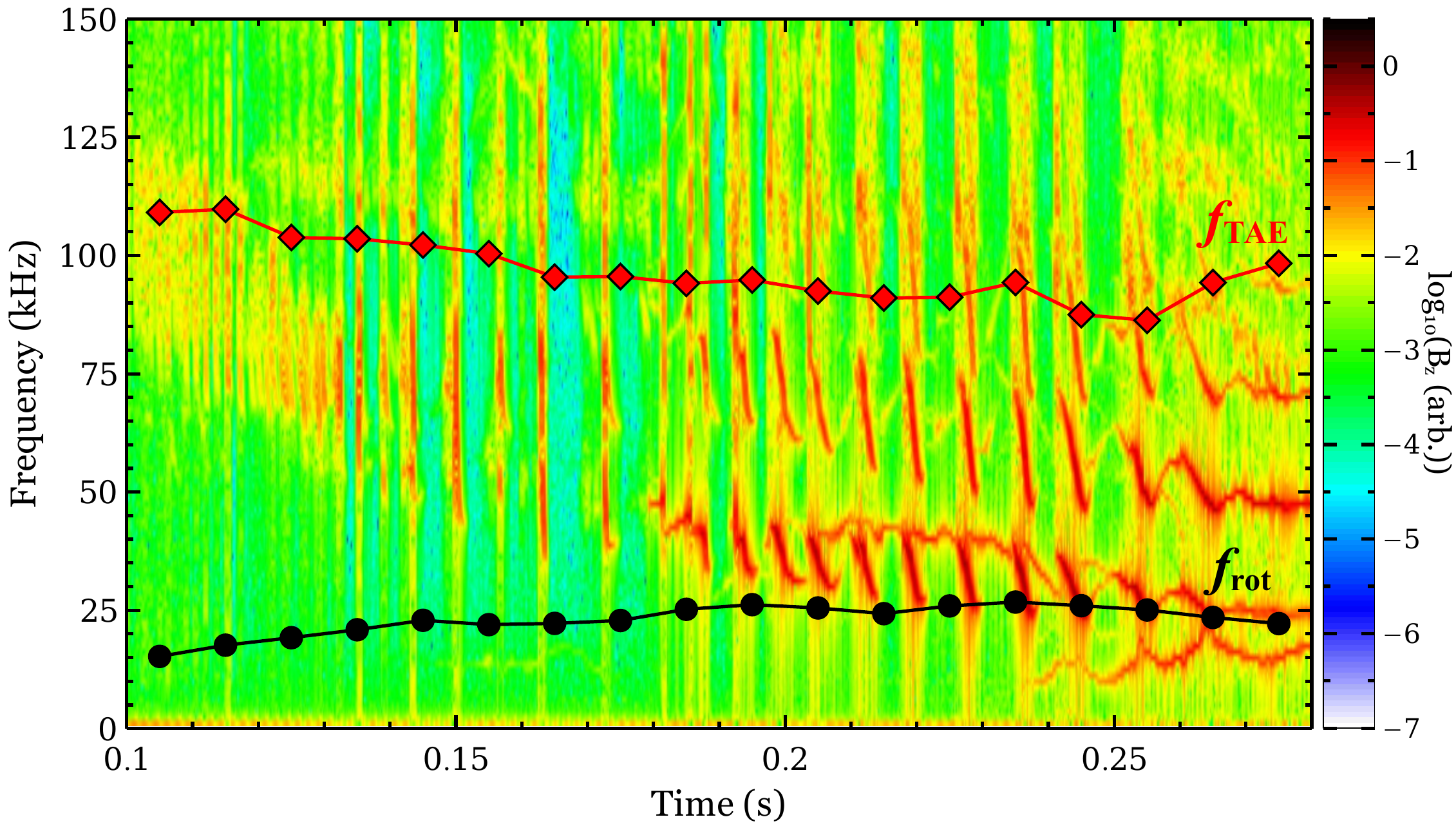}
\caption[Spectrogram of a typical MAST discharge]{The temporal evolution of the spectrum of poloidal magnetic field perturbations on the outboard midplane of the MAST vessel during shot \#29207. Black and red points show the on-axis toroidal rotation frequency and central frequency of the TAE gap respectively. The parameters used to derive these values were obtained from charge-exchange measurements of the rotation velocity, Thomson scattering measurements of the electron density, and a reconstruction of the magnetic equilibrium with the \textsc{Efit} equilibrium solver. $f_\mathrm{TAE}$ plotted here is equal to the sum of the calculated core TAE frequency and the toroidal rotation frequency, accounting for the Doppler shift of the frequency observed in the lab frame.}
\label{fig:MAST_MHD}
\end{figure}

Figure \ref{fig:MAST_MHD} shows a spectrogram of the perturbations to the poloidal magnetic field in a typical beam-heated MAST discharge. This is derived from a magnetic pick-up coil located on the outboard midplane of the vessel, close to the interior wall. Neutral beam injection at $E_b=\SI{71}{keV}$ started at \SI{0.05}{s} in this shot, injecting \SI{2.0}{MW} of NBI power. A second beam at $E_b=\SI{61}{keV}$ was added at \SI{0.18}{s}, providing an additional \SI{1.5}{MW} of NBI power.

The period shown in figure \ref{fig:MAST_MHD} exhibits closely spaced bursts of activity close to the core TAE frequency for approximately the first \SI{20}{ms}. The frequency of this activity gradually decreases over the following \SI{20}{ms}, and the bursts start to become more widely spaced in time. The low-frequency component of the well-separated bursts between \SI{0.14}{s} and \SI{0.18}{s} appears to span most of the frequency range from the core rotation frequency $f_\mathrm{rot}$ up to the TAE frequency $f_\mathrm{TAE}$; closer inspection of the signal reveals these to be \emph{chirping modes}, which sweep down rapidly in frequency over approximately \SI{1}{ms}. A clear transition occurs shortly before \SI{0.19}{s}, when the duration of each burst lengthens substantially to approximately \SI{3}{ms} and the amplitude of the bursts increases significantly. These large-amplitude bursts exhibit clear separation into a low-frequency fundamental component ($n=1$), chirping down from approximately \SI{20}{kHz} above the core rotation frequency to match $f_\mathrm{rot}$ at the end of the burst, and a second harmonic ($n=2$) at twice the fundamental frequency in the lab frame. Higher harmonics are also observed, albeit at much lower amplitudes. Finally, the bursts undergo a transition to a continuous mode which tracks the core rotation frequency from \SI{0.27}{s} until the end of the discharge.

\begin{figure}[h]
\centering
\includegraphics[width=0.85\textwidth]{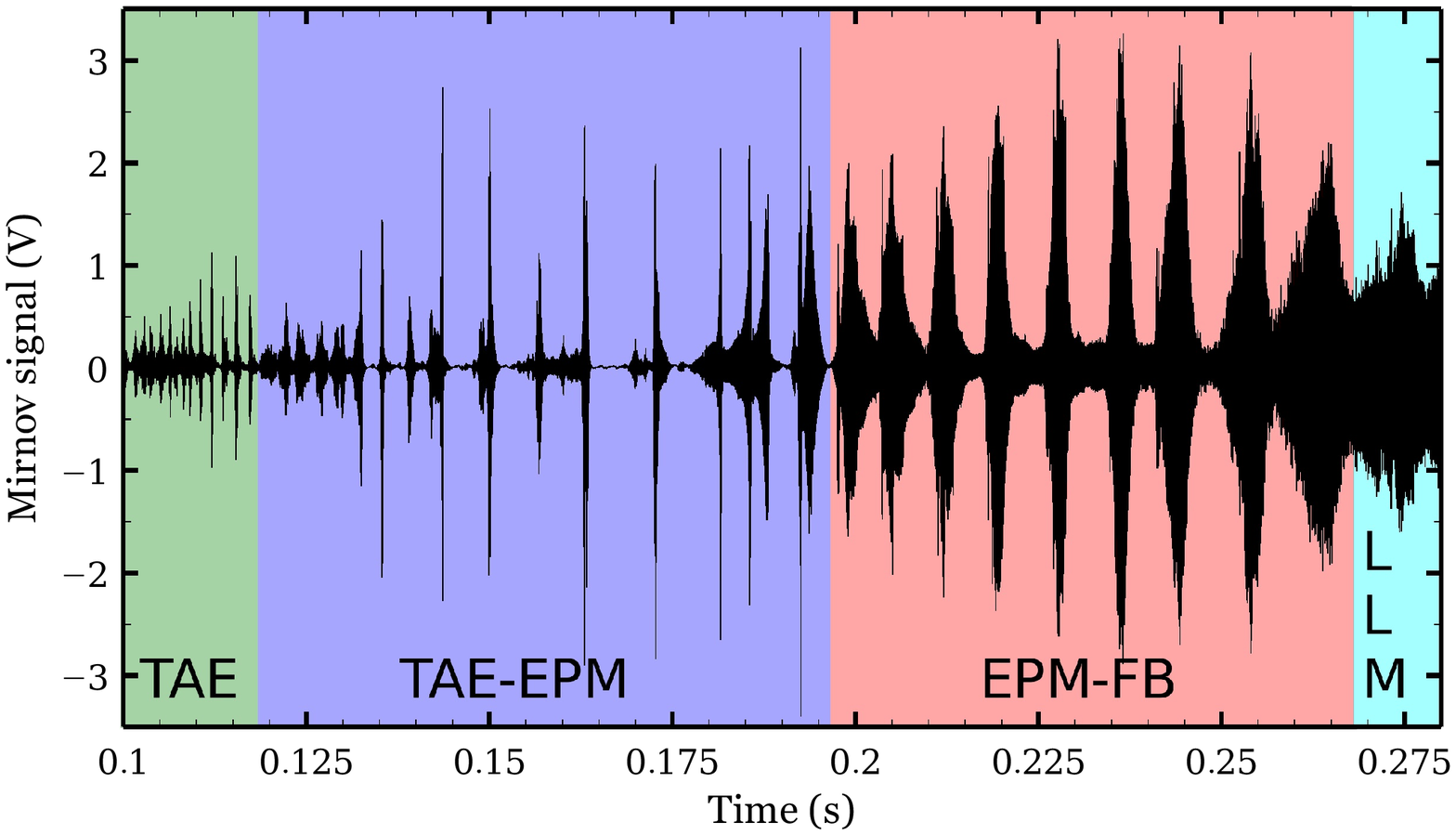}
\caption[Mirnov coil time trace during a typical MAST shot]{Time trace of the data acquired during shot \#29207 by an outboard midplane magnetic pick-up coil. The coil is the same as that used to generate the spectrogram in figure \ref{fig:MAST_MHD}. Overlaid coloured regions show the qualitative classification of the different types of MHD activity as it evolves throughout the shot. The classification is based on the temporal evolution of the frequency and amplitude of the signal. TAE$=$toroidal Alfv\'en eigenmode; EPM$=$energetic particle mode; FB$=$fishbone; LLM$=$long-lived mode, which is the name commonly given to the saturated internal kink mode in MAST.}
\label{fig:OMVtraceMHD}
\end{figure}

The evolution of the bursts of MHD activity is seen clearly in a time trace of the Mirnov pick-up coil signal. Figure \ref{fig:OMVtraceMHD} shows the raw data from the coil used to generate the spectrogram shown above.
There is no clearly defined criterion to establish on the basis of the observed behaviour whether a given burst of MHD activity should be classed as a TAE, which is a normal mode of the background plasma, or an energetic-particle mode (EPM), the existence of which is dependent on the presence of a FI population. The classification indicated in figure \ref{fig:OMVtraceMHD} is based mainly on the frequency of the mode. As seen in figure \ref{fig:MAST_MHD}, the first set of weakly chirping MHD bursts is clustered close to the core TAE frequency, so these modes are identified as TAE. The modes in the blue region in figure \ref{fig:OMVtraceMHD} exhibit much stronger frequency chirping and depart significantly from the TAE frequency, suggesting a strong energetic-particle drive which allows them to chirp down into the Alfv\'en continuum \cite{Hsu1994, Mynick1994, Breizman1997}. Features of these modes in the Mirnov coil trace include large `spikes' in amplitude lasting for just a few wave cycles; asymmetric evolution in time; highly variable spacing between bursts; and variable amplitude from one burst to the next. These are tentatively identified as TAE with nonlinear coupling to EPM, although they will be referred to simply as `chirping TAE' hereafter. In the red region, although some of the features such as the temporal asymmetry of the TAE-EPM are still apparent in the first few bursts, the evolution of each burst is slower and more regular. The amplitude of successive bursts seems to follow a slowly evolving envelope, reaching a maximum at approximately \SI{0.235}{s}, and there is much less inter-event variability. In accordance with the nomenclature used for such MHD activity on other devices \cite{McGuire1983, Heidbrink1990, Fredrickson2003, Nave2011}, these are called fishbones. A characteristic of the fishbones is that the frequency at the end of the burst coincides almost exactly with the core toroidal rotation frequency. Finally, the steady-state mode, which in MAST tends to evolve directly from the last of a set of fishbones, tracks the core rotation frequency and exhibits no frequency chirping and only weak amplitude modulation. This behaviour suggests a saturated ideal-MHD mode. Analysis with soft X-ray detectors and a toroidal array of Mirnov coils shows the dominant structure of this mode to be that of a $(m,n)=(1,1)$ internal kink \cite{Chapman2010}. This is often called the \emph{long-lived mode} (LLM) in MAST, due to its tendency to persist from the end of the fishbone period until the end of the discharge.

\begin{figure}[h]
\centering
\includegraphics[width=\textwidth]{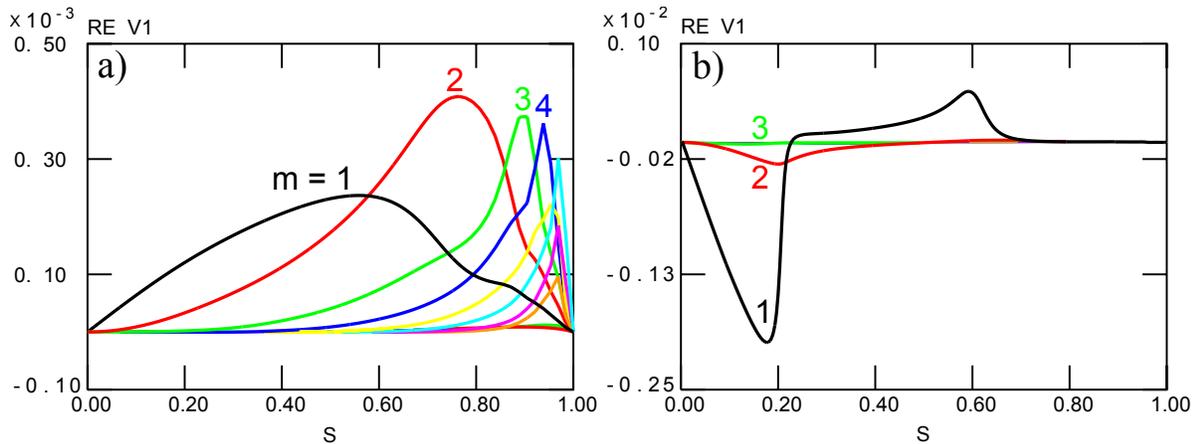}
\caption[TAE and internal kink eigenmodes in MAST]{a) Structure of the $n=1$ TAE calculated with the linear ideal-MHD code MISHKA-1 for MAST shot \#29210 at $t=\SI{0.165}{s}$. The abscissa $s$ represents the square root of normalised poloidal flux, and the ordinate represents $sv_r$, where $v_r$ is the radial velocity of the perturbation in arbitrary units. Integer labels indicate the poloidal mode number of the corresponding coloured line. b) As for panel (a), but for the $n=1$ internal kink mode in shot \#29976 at $t=\SI{0.210}{s}$.}
\label{fig:Eigenmodes}
\end{figure}

Figure \ref{fig:Eigenmodes} shows the structure of typical $n=1$ TAE and internal kink eigenmodes in MAST plasmas, calculated with the linear ideal-MHD code MISHKA-1 \cite{Mikhailovskii1997}. The broad structure of the global TAE, with multiple poloidal harmonics peaking at successively larger radii, contrasts with the predominantly $m=1$ kink mode. The kink mode exhibits a `double layer' due to the non-monotonic $q$-profile, with the internal region of negative displacement having $q\gtrsim1$ and the outer region of positive displacement having $q\lesssim1$. Two $q=1$ rational surfaces are present, at $s\approx0.2$ and $s\approx0.6$, which form the boundaries of the regions of opposite displacement. In the case of the TAE, by contrast, no $q=1$ surface exists within the plasma.

For the purposes of this study, two sets of discharges were selected for the analysis of the effects of fishbones and TAE on the FI population. Both sets consist of discharges with high NBI power and high electron density, the combination of which provides a reasonable signal-to-noise ratio (SNR) in the FI diagnostic signals along with large-amplitude bursts of TAE and fishbones separated in time by approximately $5-\SI{10}{ms}$. These points are important in allowing the effects of individual bursts of MHD activity on the FI population to be resolved by each diagnostic. A further reason for choosing these discharges is the fact that the NC was scanned in tangency radius between repeated shots, allowing composite NC count-rate profiles to be derived. Global parameters of representative discharges from each set are presented in the following section.

\section{Plasma scenarios}
\label{sec:Scenarios}
\begin{figure}[h]
\centering
\includegraphics[width=0.75\textwidth]{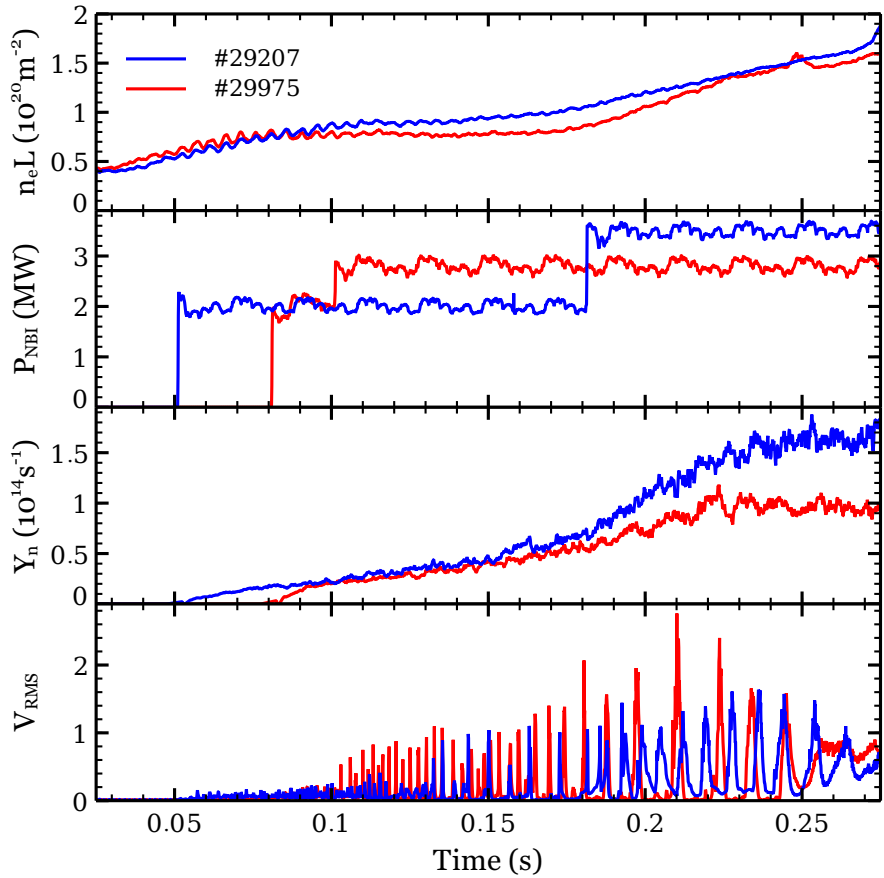}
\caption{Global parameters of representative discharges from the two sets selected for analysis of the effects of TAE and fishbones. The first set consists of shots \#29207 $-$ \#29210, and the second set includes shots \#29975, \#29976 and \#29980. The panels show, from top to bottom: line-integrated electron density measured with a CO$_2$ laser interferometer; total injected NBI power; global neutron rate measured with the fission chamber; and RMS amplitude of an outboard-midplane Mirnov coil signal sensitive to changes in the poloidal field.}
\label{fig:Plasma1}
\end{figure}

Figure \ref{fig:Plasma1} shows time traces of key parameters related to the FI population and MHD activity in a representative discharge from each of the two sets analysed in the course of this investigation.

\begin{figure}[h]
\centering
\includegraphics[width=0.75\textwidth]{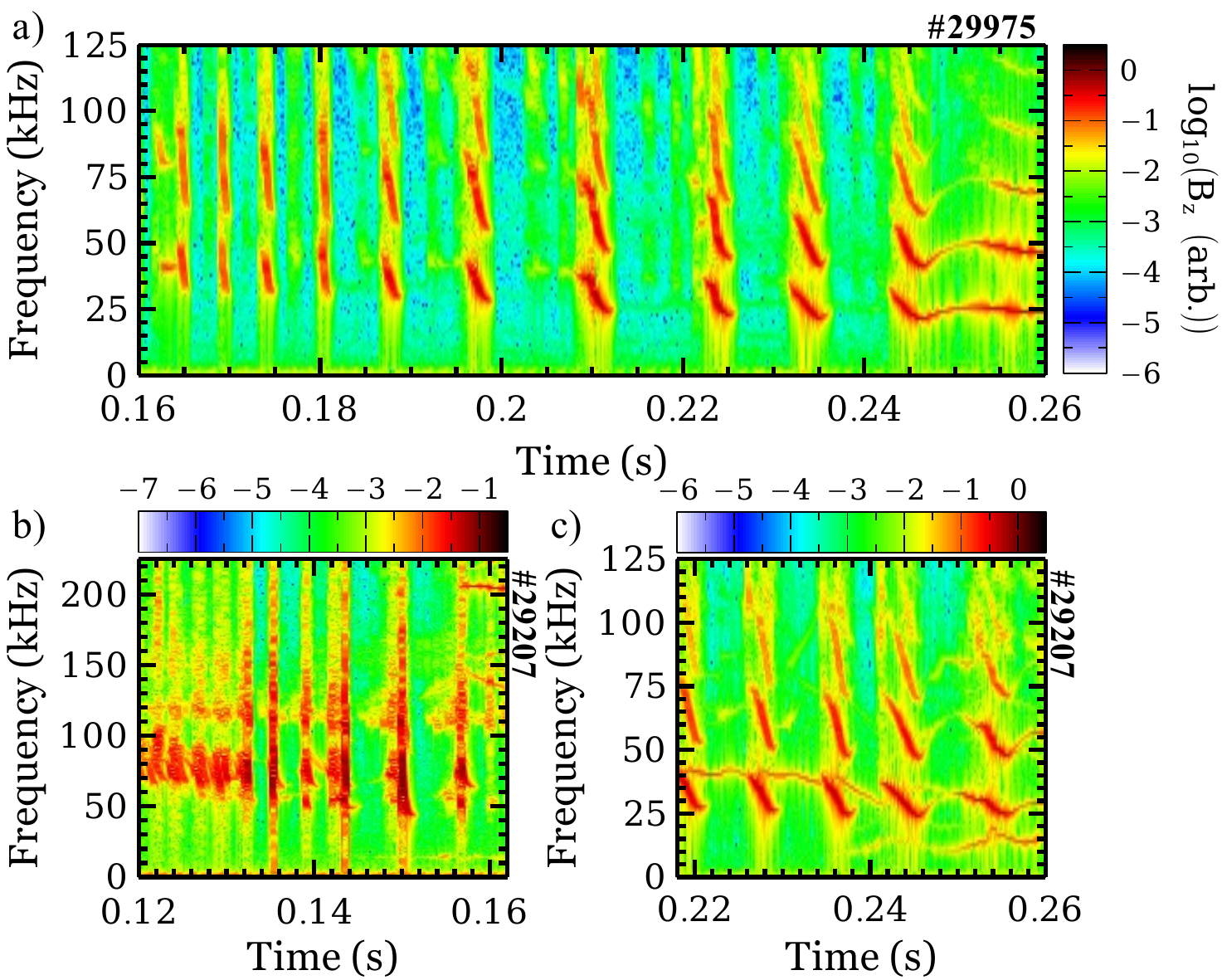}
\caption{Spectrograms from an outboard-midplane Mirnov coil signal measuring the poloidal component of the perturbed magnetic field. (a) shows a period during shot \#29975 in which chirping TAE evolve into large-amplitude fishbones. (b) and (c) show two separate periods during shot \#29207 in which chirping TAE and large-amplitude fishbones respectively dominate the low-frequency magnetic activity. Note the extended vertical axis in (b) compared to (a) and (c). In neither of these shots are there any signs of high-frequency (compressional) Alfv\'en eigenmodes, which would appear at around \SI{1}{MHz}; the absence of high-frequency MHD activity in the presence of low-frequency EPM is a characteristic feature of MAST plasmas.}
\label{fig:EPMgram}
\end{figure}

Each set of discharges featured periods with well-separated bursts of MHD activity, with the typical evolution of the magnetic spectrogram exhibited by beam-heated MAST plasmas in which chirping energetic-particle modes grow in amplitude and decrease in frequency as the $q$-profile evolves; this behaviour was illustrated in section \ref{sec:MHD}. Spectrograms from the periods of interest in each of these scenarios are shown in figure \ref{fig:EPMgram}.
Shot \#29207 \emph{et seq.} included a period with a single neutral beam injecting \SI{2.5}{MW} of power with a primary energy of \SI{71}{keV}, throughout which chirping TAE persisted, followed by a period with both neutral beams injecting a total of \SI{3.5}{MW} of power with the primary energy of the second beam set at \SI{61}{keV}, in which large quasi-periodic fishbones were observed to merge into the LLM. The second set of discharges, commencing with \#29975, exhibited very strong fishbones occurring with semi-regular spacing between bursts; the TAE are less distinct, and the early stages of the three discharges less similar to each other, than those in the first set, so the analysis of this set focuses on the large fishbones occurring late in the discharge. The total NBI power in the second set was \SI{3.0}{MW}, and the primary beam energies were \SI{70}{keV} and \SI{44}{keV}.

We now examine the data acquired with the fission chamber, neutron camera and FIDA diagnostic during each set of discharges. A systematic analysis allows firm conclusions to be drawn regarding the effects of chirping modes on the FI population.

\section{Effects of chirping modes on confined fast ions}
\label{sec:Results}
\subsection{Chirping TAE}
\begin{figure}[h]
\centering
\includegraphics[width=0.65\textwidth]{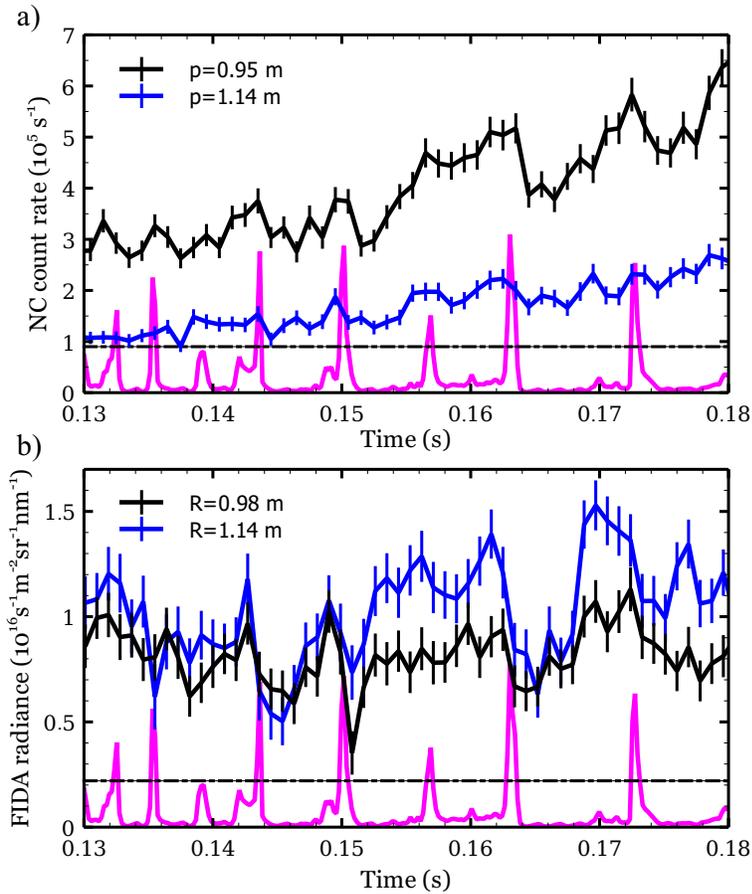}
\caption[NC and FIDA signals with chirping TAE]{a) Traces of NC count rates from the two midplane chords during a period in shot \#29207 in which chirping TAE were active. The tangency radii of the two chords are labelled, and the magenta trace shows the RMS amplitude of an outboard-midplane Mirnov coil signal in arbitrary units. b) Traces from two channels of the toroidally-viewing FIDA system during the same discharge, with the beam intersection radii labelled. Data are averaged over wavelengths \SIrange{660.7}{661.3}{nm}, and therefore include FIDA emission from passing FI with energy $E>\SI{46}{keV}$. Error bars on each trace represent random uncertainty.}
\label{fig:FIDtraceTAE}
\end{figure}

As shown in figure \ref{fig:FIDtraceTAE}, a significant reduction in the NC count rate and FIDA radiance is observed to coincide with many TAE bursts in the first set of discharges.
The timing of these drops in signal relative to the amplitude of the envelope of the Mirnov coil signal varies substantially between events. Not all drops in the diagnostic signals are associated with prominent chirping modes, neither are all bursts of magnetic activity correlated with significant changes in the diagnostic signals. In light of this variability, a quantitative approach must be adopted to determine the degree to which the reduction in FI confinement indicated by the NC and FIDA signals is correlated with chirping TAE. No CFPD data are available for this set of discharges, since the diagnostic had not been installed at the time.

The first stage in identifying the correlation between TAE and FI diagnostics signals is to establish criteria for the identification of an `event' in each case. Inspection of time traces such as those shown in figure \ref{fig:FIDtraceTAE} suggests that strong effects on the diagnostic signals are associated with TAE bursts with a large amplitude in the RMS Mirnov coil trace. Any increase in magnetic activity which exceeds a threshold RMS amplitude is therefore chosen to constitute an `event' in the magnetics signal; this threshold is indicated by the dot-dashed horizontal lines in figure \ref{fig:FIDtraceTAE}.
\begin{figure}[h]
\centering
\includegraphics[width=0.6\textwidth]{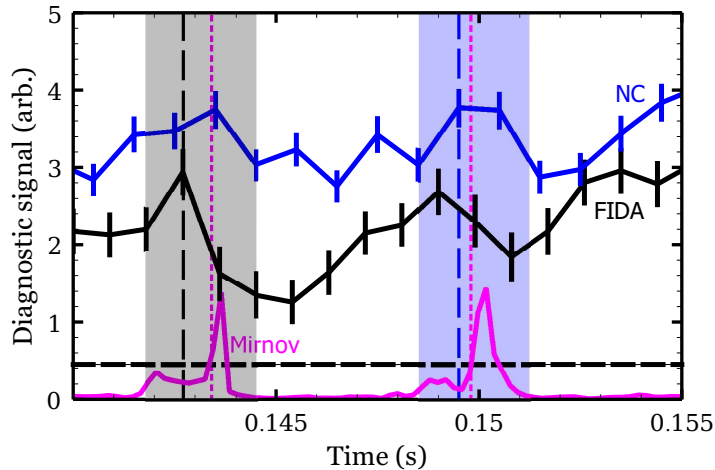}
\caption[Identification of events in FI diagnostic signals]{Drops in NC and FIDA signals which qualify as `events' for the purposes of determining the correlation with chirping TAE. The dashed vertical black and blue lines mark the onset times of events in the FIDA and NC signals respectively. Dotted vertical magenta lines mark the onset times of events in the RMS Mirnov coil signal, in which the magenta trace exceeds the threshold shown by the dashed horizontal line. The shaded regions which extend from one time step before to two time steps after the events in FI diagnostic signals indicate the period within which the magnetics event must occur in order to be counted towards successful identification; clearly both of the events highlighted here satisfy this criterion.}
\label{fig:EvID}
\end{figure}
In order to exclude drops in the NC and FIDA signals which occur due to statistical noise, or spurious drops in the nett FIDA signal caused by rapid, transient spikes in background emission correlated with TAE bursts, the condition imposed on these signals is that an `event' constitutes a drop in signal which does not recover to the pre-event level within four time steps. With the integration time of the FIDA data set at \SI{0.9}{ms} and that of the NC data set at \SI{1.0}{ms}, this provides sufficient temporal resolution to allow each of the TAE bursts to be separated while ensuring that most of the `events' represent true reductions in the part of the FI population to which the diagnostics are sensitive. Examples of qualifying events in the NC and FIDA signals are shown in figure \ref{fig:EvID}. Note that a quadratic trend is removed from each of the time traces prior to analysis; this ensures that significant drops in diagnostic signals are not masked by the trend of the signal throughout the analysis period, which extends from \SIrange{0.13}{0.18}{s} in each of the four discharges analysed in this work.

\begin{figure}[ht]
\centering
\includegraphics[width=0.6\textwidth]{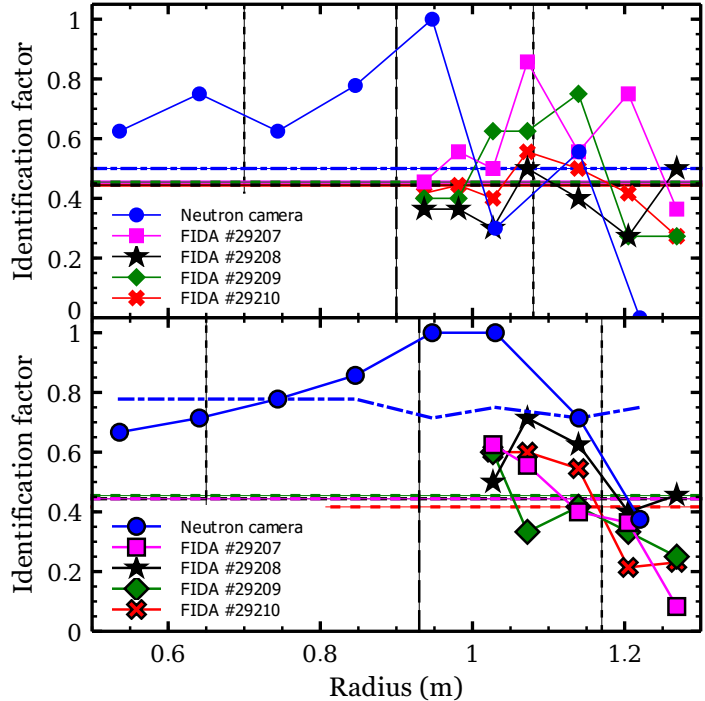}
\caption[TAE and fishbone identification factors]{Top panel: TAE identification factors, defined according to equation \ref{eq:QID}, for the NC and FIDA diagnostics during the period \SIrange{0.13}{0.18}{s} in shots \#29207 $-$ \#29210. Vertical dashed and dotted lines mark the approximate positions of the magnetic axis and $q_\mathrm{min}$ surfaces respectively. Dashed horizontal lines of various colours mark the 90\% confidence level for event identification in the corresponding FI diagnostic signal; data points have less than a 10\% chance of lying above these lines assuming random time series. Bottom panel: as above but for fishbones occurring in the period \SIrange{0.208}{0.258}{s} during the same set of discharges. The two innermost FIDA channels are excluded from this panel due to the contamination of the spectra by SW beam emission.}
\label{fig:Qfac}
\end{figure}

Proceeding from the strict definition of an `event' in the magnetics and FI diagnostic time traces, the criterion chosen to determine whether events coincide is that the onset time of the magnetics event must lie within a period extending from one time step before to two time steps after the start of the drop in FIDA or NC signal. This period is indicated by the shaded regions in figure \ref{fig:EvID}. The natural variation between events, arising due to the variable levels of drive and damping of the TAE, is thereby taken into account. For each FI diagnostic time trace, and for the analysis period during each discharge, an `identification factor' may be defined to quantify the correlation between events in the diagnostic signals and the occurrence of TAE bursts. This factor is defined as
\begin{equation}
Q_\mathrm{ID}=\frac{\mathrm{identified\:events}}
					{\mathrm{total\:TAE\:events}+\mathrm{false\:positives}}.
\label{eq:QID}
\end{equation}
The top panel of figure \ref{fig:Qfac} shows the results of the TAE identification analysis for all four discharges in this set. In each case, the FI diagnostics are best able to identify the effects of TAE bursts in the radial range where the signal is strongest; in the case of the NC this corresponds to the region close to the magnetic axis, and in the case of the FIDA diagnostic this corresponds to mid-radius on the outboard side. This observation suggests that the limiting factor in the ability of the diagnostics to resolve the effects of chirping TAE on FI, at least in this set of discharges, is the SNR.

A meaningful assessment of the significance of these identification factors requires confidence intervals to be determined. The process adopted here was to randomly permute the data points in each time series, and to establish an identification factor $Q_{\mathrm{ID}}^{*}$ based on the new, randomised time series. Repeating this process many times allows confidence levels to be quantified as percentiles of all $Q_{\mathrm{ID}}^{*}$. If the $Q_\mathrm{ID}$ of the original time series lies above, say, the $n\%$ confidence level, then the probability of the observed value of $Q_\mathrm{ID}$ arising purely by chance is less than $(100-n)\%$. Dashed, horizontal lines in figure \ref{fig:EvID} mark the 90$^\mathrm{th}$ percentile of 10,000 values of $Q_{\mathrm{ID}}^{*}$, which is to say the 90\% confidence level. The advantage of this procedure is that it is insensitive to the distribution of data points in the time series as long as the properties of this distribution are constant throughout the analysis period. To ensure that this was the case, the time series (with least-squares fitted quadratic trends removed) were subjected to the Dickey-Fuller test for the presence of unit root \cite{Dickey1979}. In all cases, the null hypothesis (that a unit root is present in the time series) was rejected at the 1\% level, allowing the time series to be treated as stationary. It can therefore be said with confidence that chirping TAE cause a significant reduction of the confined FI density in these plasmas.

\subsection{Fishbones}
Each of the discharges in the set analysed above also exhibited well-spaced, quasi-periodic fishbones as seen in panel (c) of figure \ref{fig:EPMgram}. The method described in the previous subsection was applied to the NC and FIDA signals to establish the correlation between these signals and the fishbones. In this analysis, the threshold applied to the RMS Mirnov coil signal was increased to account for the larger amplitude of the fishbone bursts compared to the TAE bursts.
\begin{figure}[h]
\centering
\includegraphics[width=0.65\textwidth]{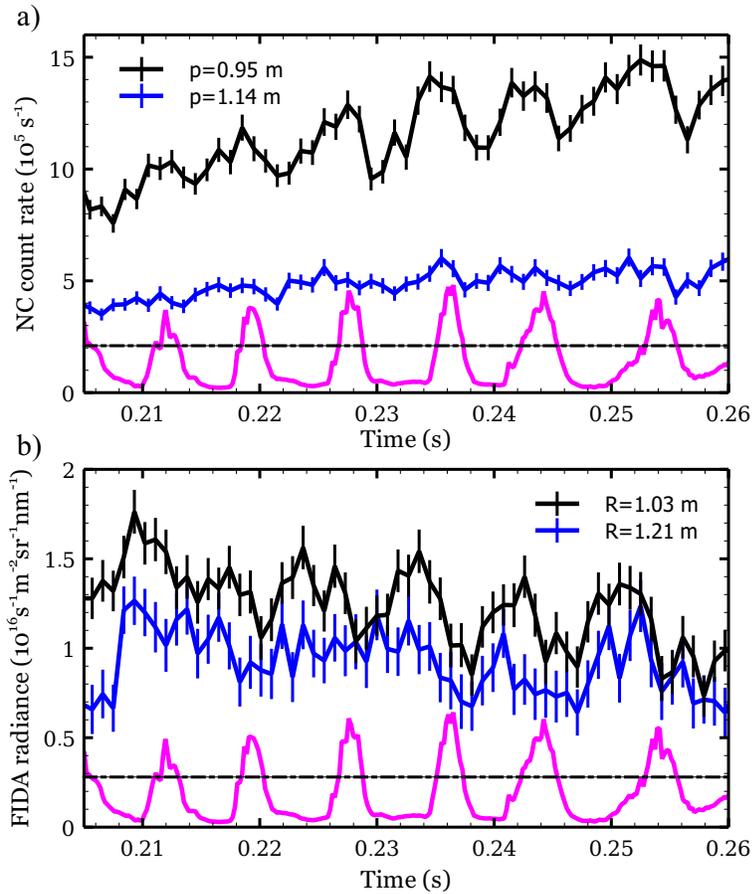}
\caption[NC and FIDA signals with fishbones]{a) Traces of NC count rates from the two midplane chords during a period in shot \#29207 in which fishbones were active. The tangency radii of the two chords are labelled, and the magenta trace shows the RMS amplitude of an outboard-midplane Mirnov coil signal in arbitrary units. b) Traces from two channels of the toroidally-viewing FIDA system during the same discharge, with the beam intersection radii labelled. Data are averaged over wavelengths \SIrange{660.7}{661.3}{nm}, and therefore include FIDA emission from passing FI with energy $E>\SI{46}{keV}$. Error bars on each trace represent random uncertainty.}
\label{fig:FIDtraceFish}
\end{figure}
Dashed, horizontal lines in each panel of figure \ref{fig:FIDtraceFish} indicate the chosen threshold. The width of the time window which was searched for a burst in Mirnov coil signal corresponding to each drop in NC or FIDA signal was also widened compared to that used in the TAE analysis. This widening accounts for the fact that the fishbones evolve more slowly than the TAE and therefore take longer to have a significant effect on the FI population. The window was extended by half a time step in each direction, thus covering the period from 1.5 time steps before to 2.5 time steps after the commencement of the drop in FI diagnostic signal. Significant drops in NC and FIDA signals are clearly seen to accompany the fishbone bursts in figure \ref{fig:FIDtraceFish}. The results of the analysis described above are shown in the bottom panel of figure \ref{fig:Qfac}. As in the case of the TAE, both diagnostics see significant drops in signal correlated with the fishbones. In many of the channels the confidence in the correlation exceeds 90\%. Note that the two innermost channels of the FIDA system cannot be included in this analysis since the spectra are strongly contaminated with SW beam emission.

In previous studies of MAST FIDA data \cite{Jones2013}, it was found that the relative change in core FIDA signal due to the fishbones correlated strongly with the maximum RMS amplitude of the Mirnov coil signal and only weakly with the amplitude of the perturbation. This analysis was repeated for the data from the recent set of discharges (\#29207 \emph{et seq.}).
\begin{figure}[h]
\centering
\includegraphics[width=0.95\textwidth]{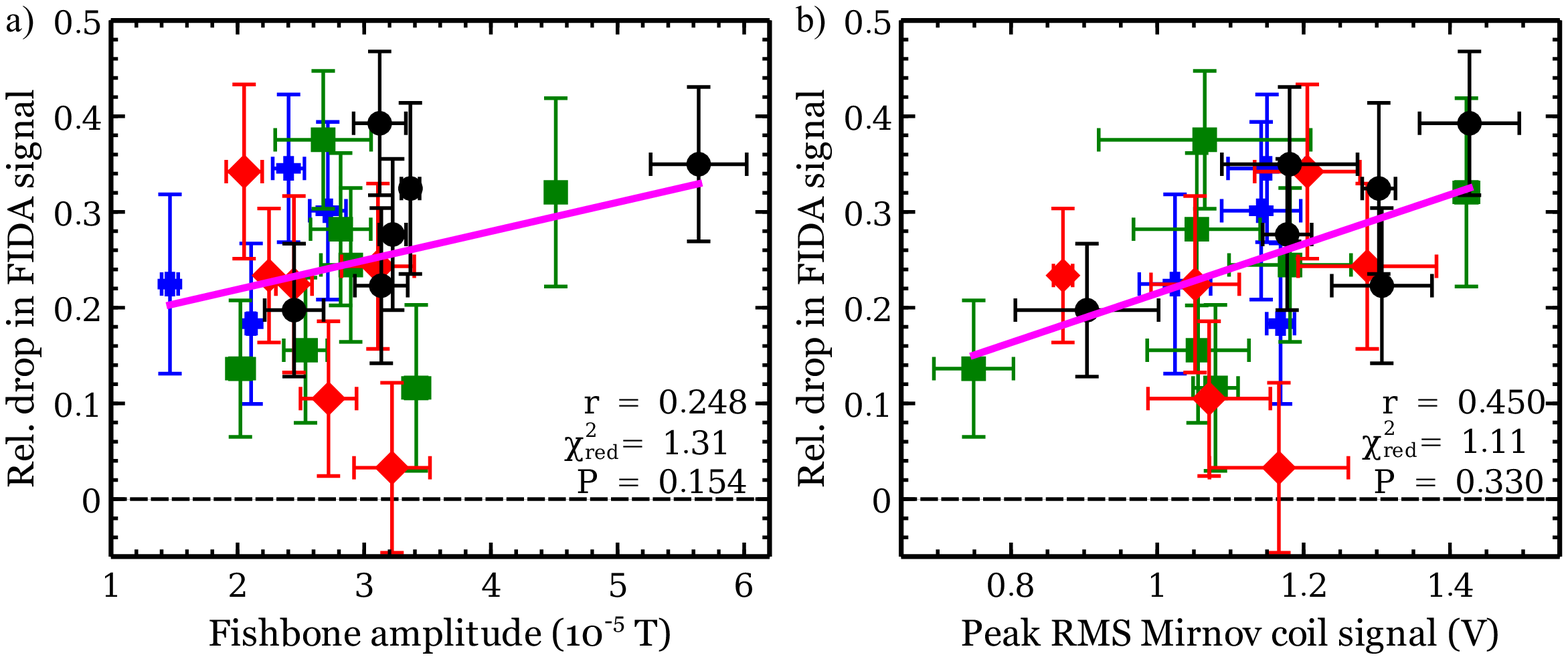}
\caption[Drop in FIDA signal scales with fishbone amplitude]{Relative change in FIDA signal plotted against (a) the amplitude of the fishbone perturbation and (b) the maximum RMS amplitude of the Mirnov coil signal for a set of 23 fishbones in shots \#29207 $-$ \#29210. Each data point corresponds to a single fishbone, and different symbols are used for each discharge. The relative drop in signal is defined as $(S_i-S_f)/S_i$ where $S_i$ and $S_f$ are the signal before and after the fishbone respectively, The correlation coefficient $r$, reduced $\chi^2$ and cumulative chi-square probability $P=\Pr(X^2\geq\chi^2)$ are shown in each panel. Data are averaged over $R=$ \SIrange{1.03}{1.14}{m} at $\lambda=\SI{660.7}{nm}$.}
\label{fig:magcorr}
\end{figure}
Figure \ref{fig:magcorr} shows the scaling of the drops in FIDA signal with the amplitude of the fishbones and with the RMS amplitude of the perturbation in the Mirnov coil signal. The wavelength used in figure \ref{fig:magcorr} corresponds to a minimum FI energy of \SI{46}{keV}. The data are averaged over three channels with beam intersection radii \SI{1.03}{m}, \SI{1.07}{m} and \SI{1.14}{m}. A different symbol is used for each shot, but the regression analysis takes into account the whole data set. As in the earlier study \cite{Jones2013}, the drops in FIDA signal are correlated much more strongly with the RMS Mirnov coil signal than with the fishbone amplitude. A similar linear scaling was found between the drops in global neutron rate and the peak Mirnov coil signal in a previous study of MAST data \cite{Turnyanskiy2013}. This suggests that the strength of the interaction between the fishbones and the FI is sensitive to the frequency of the perturbation as well as to its amplitude, since $\dot{B}_\theta\approx\omega\tilde{B}_\theta$. Here, $\dot{B}_\theta$ is the time derivative of the poloidal magnetic field, to which the Mirnov coil signal is proportional assuming a fixed mode location, and $\omega$ and $\tilde{B}_\theta$ are the frequency and amplitude of the mode. Such a relationship was previously found between drops in the global neutron rate and $\langle\dot{B}_\theta\rangle_\mathrm{RMS}$ during TAE bursts in DIII-D \cite{Duong1993a}, while no linear dependence on $\langle\dot{B}_\theta\rangle_\mathrm{RMS}$ was seen in the case of fishbone bursts. A linear correlation was later found, albeit in a smaller data set, between drops in the neutron rate and the amplitude $\tilde{B}_\theta$ of chirping modes in the fishbone-EPM frequency range \cite{Heidbrink1999a}. In NSTX meanwhile, a very weak linear scaling was identified between FI losses and mode amplitude $\tilde{B}_\theta/B$ in the case of both EPM and TAE, with TAE having a stronger effect for a given mode amplitude \cite{Fredrickson2006}. Fishbones were not included in that data set, and the correlation with $\dot{B}_\theta$ was not examined. For the changes in FIDA signal associated with the chirping TAE discussed in the previous subsection, no meaningful correlation was found with either $\langle\dot{B}_\theta\rangle_\mathrm{RMS}$ or with $\tilde{B}_\theta$. The present result therefore raises the question of which quantity is the better proxy for the ability of chirping modes to redistribute FI, particularly in the presence of mixed low and high-frequency modes.
 
Since the tangency radius of the NC was scanned between discharges in this set, the correlation analysis presented in figure \ref{fig:magcorr} unfortunately cannot be extended to this diagnostic. Furthermore, the drops in global neutron rate due to these fishbones are small and difficult to distinguish from noise or from perturbations caused by changes in bulk plasma parameters or non-resonant MHD modes. We therefore move on to consideration of the possibility that as well as causing real-space or velocity-space transport of resonant FI as suggested by the results in this section, the TAE and fishbones may also cause losses of FI from the plasma.

\subsection{Fast-ion losses due to resonant MHD instabilities}
\begin{figure}[h]
\centering
\includegraphics[width=0.65\textwidth]{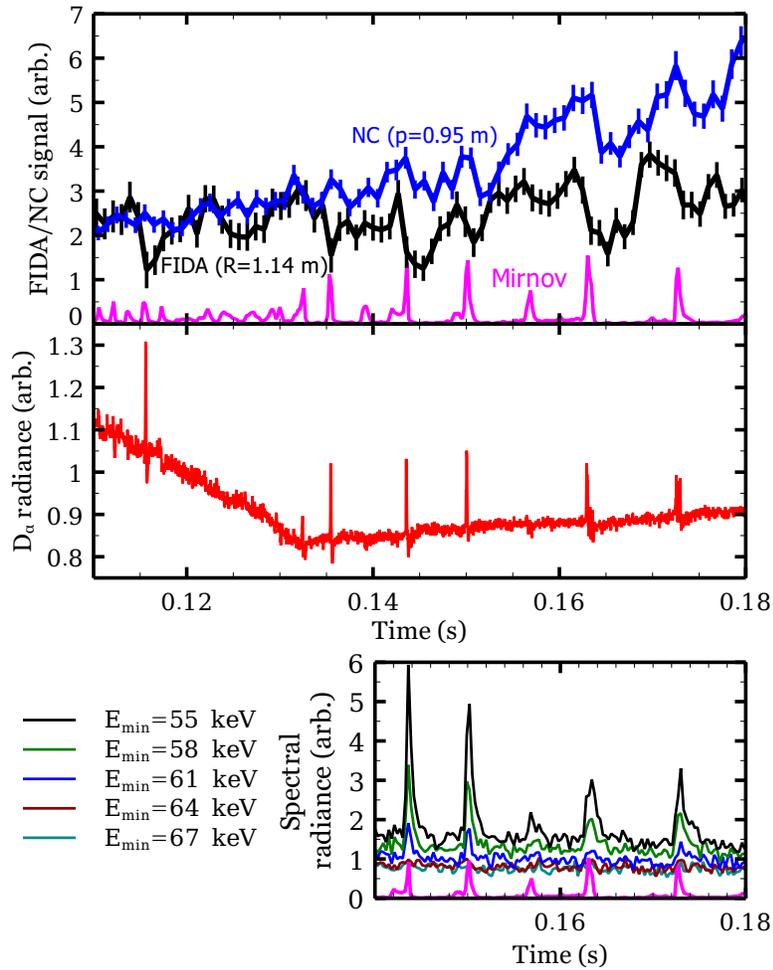}
\caption[D$_\alpha$ spikes coincide with chirping TAE]{Chirping TAE during shot \#29207 are observed to cause large drops in core NC and FIDA signals which in most cases take several milliseconds to recover to the pre-event level (top panel). Coincident with the largest of these bursts, spikes in D$_\alpha$ radiation from the edge of the plasma are observed (middle panel). A large drop in FIDA signal and a spike in edge D$_\alpha$ are also observed at around \SI{0.115}{s}, despite the fact that only a small TAE, and no drop in NC signal, is observed at this time. The nature of this event is unknown, but the observations suggest a spatially-localised loss of FI. The bottom-right panel shows the signal from a background-viewing FIDA channel at wavelengths corresponding to minimum detectable FI energies $E_\mathrm{min}$ close to the beam injection energy, which was \SI{71}{keV}. This signal comprises bremsstrahlung and passive FIDA emission, as discussed in the text. The midplane intersection radius of the LOS from which these data are obtained is $R=\SI{1.33}{m}$.}
\label{fig:DaTAE}
\end{figure}

\begin{figure}[h]
\centering
\includegraphics[width=0.65\textwidth]{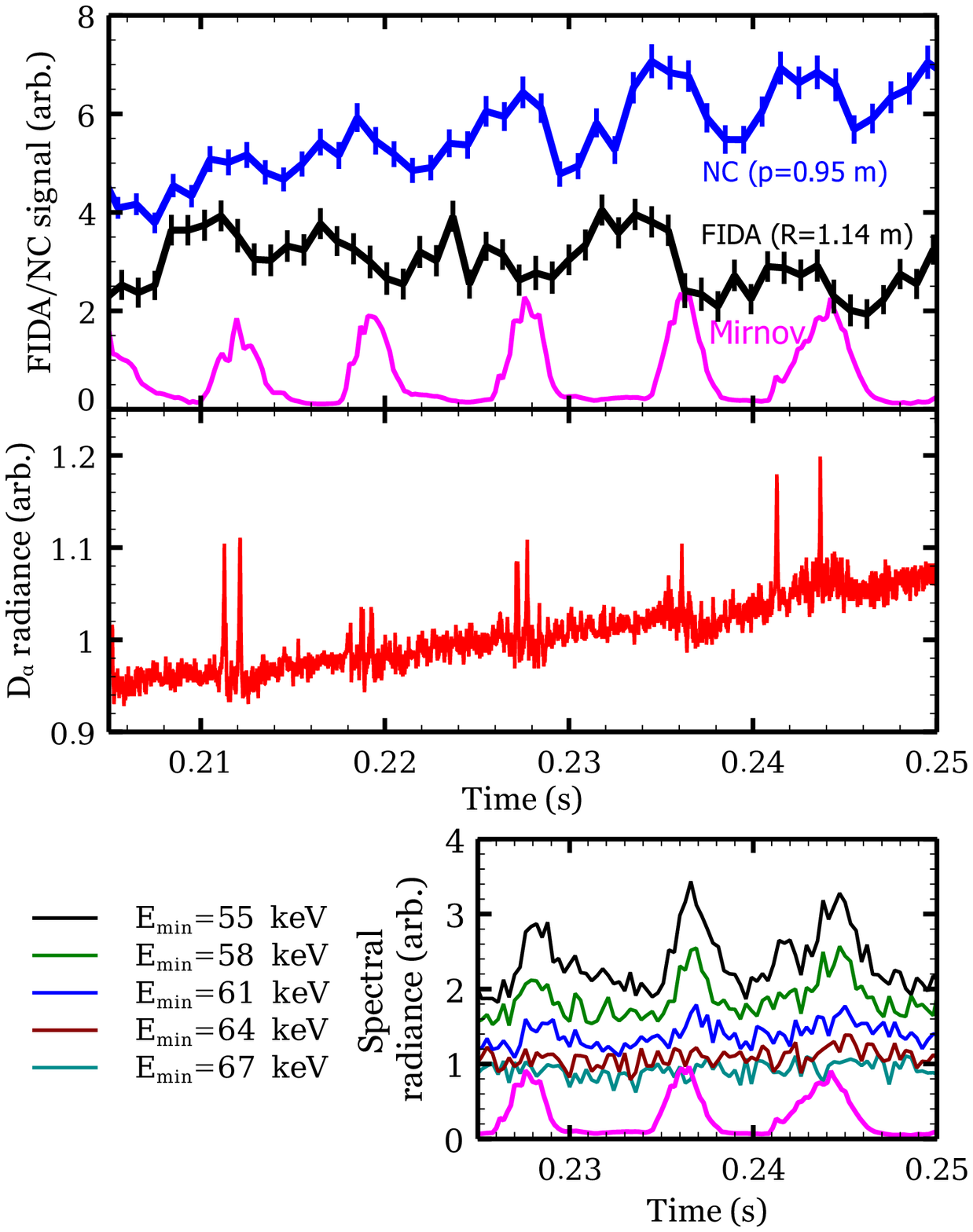}
\caption[D$_\alpha$ spikes coincide with fishbones]{Fishbones during shot \#29207 are observed to cause large drops in core NC and FIDA signals which in most cases take several milliseconds to recover to the pre-event level (top panel). Coincident with these bursts, spikes in D$_\alpha$ radiation from the edge of the plasma are observed (bottom panel). The D$_\alpha$ spikes in this case exhibit more structure, last longer, but have a smaller amplitude than in the case of the TAE. As in the previous figure, the bottom right panel shows background emission (bremsstrahlung and passive FIDA) at wavelengths corresponding to minimum detectable FI energies $E_\mathrm{min}$ close to the beam injection energy.}
\label{fig:DaFish}
\end{figure}

Spikes in the D$_\alpha$ radiation from the edge of the plasma and from the divertor are observed to coincide with many chirping modes. Figures \ref{fig:DaTAE} and \ref{fig:DaFish} show examples of this behaviour from shot \#29207.
Bursts in the Mirnov coil trace coincide with a significant reduction in FIDA and neutron emission from the core of the plasma, as well as with pronounced spikes in the signal from a filtered D$_\alpha$ monitor viewing the edge of the plasma close to the midplane. These spikes in D$_\alpha$ light were hypothesised to be the result of CX reactions between FI ejected from the plasma by the modes, and neutrals close to the plasma boundary. The open divertor in MAST meant that deuterium recycled from the plasma at the divertor plates was able to circulate freely in the vacuum vessel, forming a dense `blanket' of neutrals at the edge of the plasma.

To test this hypothesis regarding the origin of the D$_\alpha$ spikes, the \emph{passive FIDA} emission provides useful information. Passive FIDA is the D$_\alpha$ light emitted by reneutralised FI which have undergone CX with edge neutrals rather than with beam or halo neutrals. By inspecting the time trace of passive FIDA during a time window in which edge D$_\alpha$ spikes are observed, the mechanism giving rise to these spikes is firmly established to be related to the FI. The bottom panels of figures \ref{fig:DaTAE} and \ref{fig:DaFish} show time traces of passive FIDA from a reference channel ($R=\SI{1.33}{m}$) during the period in which large-amplitude TAE or fishbones and D$_\alpha$ spikes are observed. The spikes in passive FIDA emission are clearly seen to be suppressed close to the beam injection energy, which is \SI{71}{keV} during this shot. This observation demonstrates conclusively that the D$_\alpha$ spikes are at least partially caused by reneutralised FI rather than by thermal neutrals or perturbations to bremsstrahlung radiation. This implies that both TAE and fishbones caused losses of FI from these plasmas.

Analysis and interpretation of the FIDA and NC observations reveals that both TAE and fishbones cause FI to be transported from the core of the plasma and even lost through the plasma boundary. The next step in this analysis is to attempt to model the effects of the modes on FI using the \emph{ad hoc} models available within \textsc{Transp}.

\section{TRANSP modelling of fast-ion redistribution}
\label{sec:Transp}
\subsection{Anomalous transport models}
Two mechanisms are available within \textsc{Nubeam} to allow the effects of TAE and fishbones on the FI to be modelled. The redistribution may be modelled as a combination of diffusive and convective transport, with user-specified diffusion and advection coefficients specified as a function of radial position, time and energy. Without first-principles modelling of the underlying physics to guide the choices of these coefficients however, only the simplest models are practically applicable. In the modelling described in this section, anomalous diffusion was varied as a function of time but was taken to be radially uniform throughout the plasma. Convective transport was not applied, since diffusive transport was found to reproduce the observations reasonably well. The effect of diffusion is to cause radial transport of FI at a rate proportional to the radial gradient of the FI density. The resulting flux is
\begin{equation*}
\Gamma_\mathrm{FI} = -D_\mathrm{an} \nabla_r n_\mathrm{FI},
\end{equation*}
where $D_\mathrm{an}$ is the spatially and temporally-varying anomalous FI diffusivity. The diffusive model was applied successfully to model the global neutron rate and NC count rate profiles in MAST discharges with on- and off-axis NBI \cite{Turnyanskiy2013}.

The second FI transport mechanism available within \textsc{Nubeam} is an \emph{ad hoc} fishbone loss model. This model takes as input parameters: the time of the first fishbone and the time at which the model is switched off; the duration of the fishbone burst and the characteristic loss time of the FI while the fishbone is active; the time between fishbone bursts; and the minimum and maximum values of energy and pitch of the affected FI. This model was successfully used to reproduce global and local neutron emission from MAST plasmas as described by Klimek \emph{et al.} \cite{Klimek2015}. In the present work however, FIDA observations are used as an additional constraint on the parameters of the model.

In the following subsections the results of \textsc{Transp} modelling using anomalous diffusion and the fishbone model are presented.

\subsection{Modelling with anomalous diffusion}
\begin{figure}[h]
\centering
\includegraphics[width=0.75\textwidth]{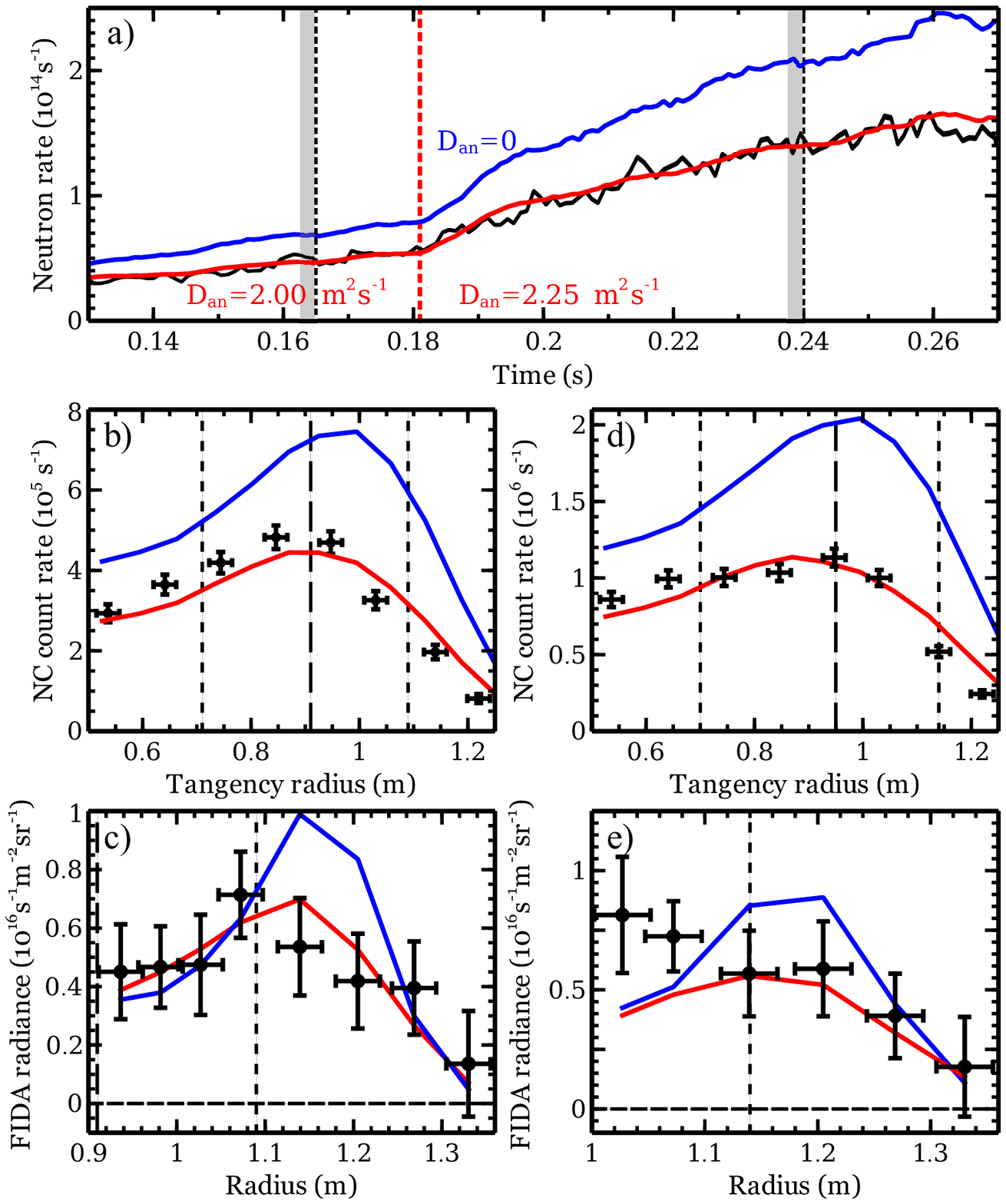}
\caption[\textsc{Transp} modelling with anomalous diffusion]{Comparisons between measured and \textsc{Nubeam}-modelled FI diagnostic time traces and radial profiles in the presence of chirping TAE and fishbones. Panel (a) shows the global neutron rate measured in shot \#29210 (black) compared to that modelled with (red) and without (blue) spatially-uniform anomalous FI diffusion in \textsc{Transp}. The values of anomalous diffusivity used in these simulations are given in the labels. The forward-modelled NC and FIDA profiles are compared to the measured profiles in panels (b) and (c) for $t=\SI{0.165}{s}$ and in panels (d) and (e) for $t=\SI{0.240}{s}$. Contamination of the core FIDA signal with SW beam emission at the later time means that the two channels between \SI{0.9}{m} and \SI{1.0}{m} cannot be included, and the signal in the two channels between \SI{1.0}{m} and \SI{1.1}{m} is artificially elevated. Vertical dashed and dotted lines in panels (b) to (e) indicate the approximate radial positions of the magnetic axis and minimum-$q$ surfaces. $\lambda=[660.9-661.9]$ nm for FIDA data.}
\label{fig:29207serCmp}
\end{figure}

The series of discharges \#29207 \emph{et seq.}, with chirping TAE and quasi-periodic fishbones, provided an opportunity to test the diffusive \textsc{Nubeam} modelling in the presence of resonant MHD activity. Time slices averaged over \SI{3}{ms}, from \SIrange{0.162}{0.165}{s} during the chirping TAE phase and from \SIrange{0.237}{0.240}{s} during the fishbone phase, were used for a comparison between \textsc{Nubeam} modelling and FI diagnostic signals under the assumption of spatially-uniform anomalous diffusion. The results of this comparison are shown in figure \ref{fig:29207serCmp}.

Clearly the introduction of anomalous diffusion is sufficient in this case to model the effects of both chirping TAE and fishbones on the FI distribution, at least to the extent that the distribution may be constrained by the available diagnostics. It is worth noting that at the two times shown in figure \ref{fig:29207serCmp} the measured profiles are averaged over periods during which a significant MHD perturbation occurred; profiles during selected MHD-quiescent windows shortly before a TAE or fishbone burst exhibit increased signal levels, although the signal still does not reach the level modelled in the absence of anomalous FI diffusion. The coarse assumption of spatially-uniform diffusivity, held constant over a long period, cannot capture the dynamics of individual bursts of MHD and the interaction of these bursts with the FI.

\subsection{Inclusion of the fishbone model}
\begin{figure}[h]
\centering
\includegraphics[width=\textwidth]{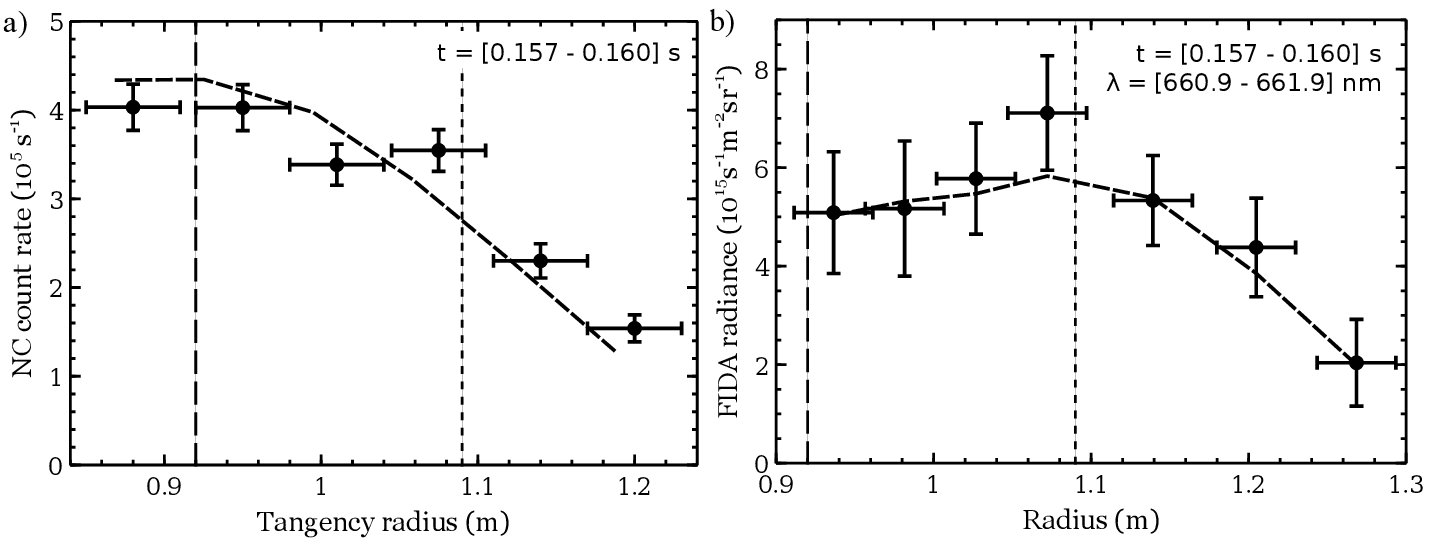}
\caption[Modelling \#29976 with anomalous diffusion]{Profiles of measured and modelled NC and FIDA signals averaged over the period \SIrange{0.157}{0.160}{s} during shot \#29976 (with the measured NC profile including data from similar discharges \#29975 and \#29980). The modelled profile is based on a simulation with an anomalous diffusivity of \SI{3.0}{m^2.s^{-1}} at this time during the discharge. Vertical lines mark the approximate radial positions of the magnetic axis and minimum-$q$ surface.}
\label{fig:29975serEarly}
\end{figure}

Figure \ref{fig:29975serEarly} shows that spatially-uniform anomalous diffusion models well the early part of shot \#29975 \emph{et seq.} during which chirping TAE were active. The forward-modelled NC and FIDA profiles show good agreement with the measured profiles at $t=\SI{0.160}{s}$, when a diffusivity of $D_\mathrm{an}=\SI{3.0}{m^2.s^{-1}}$ allows the \textsc{Transp} modelling to match the measured global neutron rate to within 10\%.

\begin{figure}
\centering
\includegraphics[width=0.6\textwidth]{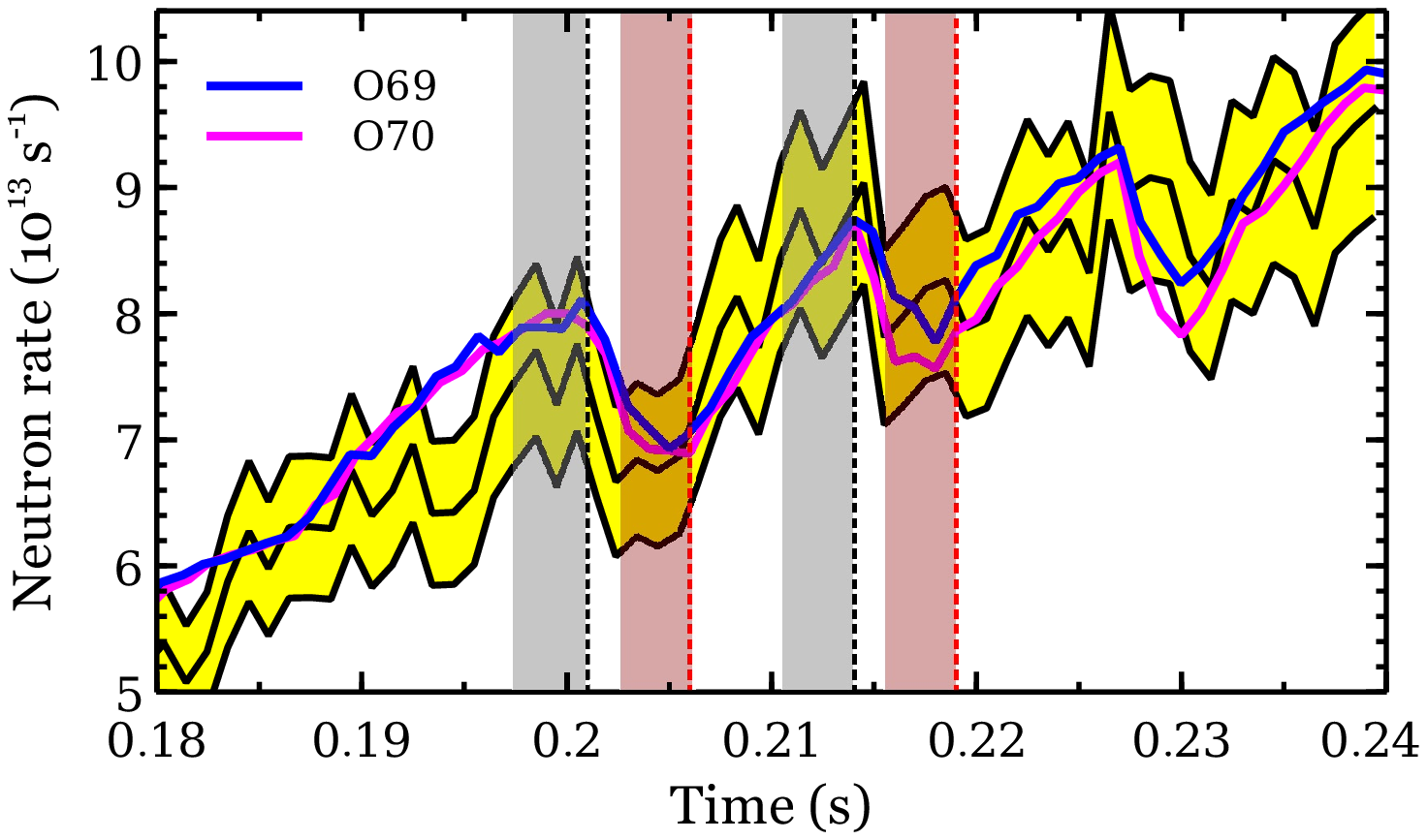}
\caption[Neutron rate match with \textsc{Transp} fishbone model]{A comparison between the measured (black trace, with yellow shaded region indicating statistical uncertainty) and modelled (blue and magenta traces) global neutron rates for shot \#29976 with the fishbone model applied within \textsc{Nubeam}. The fishbones were specified to occur from \SI{0.201}{s} to \SI{0.230}{s} with a period of \SI{12.8}{ms} and a duration of \SI{4}{ms}. The characteristic loss time of the affected FI was \SI{3}{ms} in run O69 and \SI{1}{ms} in run O70. In run O69, FI with $\SI{50}{keV} \leq E \leq \SI{75}{keV}$ and $0.69 \leq p \leq 0.93$ (evaluated at either of the two previous midplane crossings) were removed by the fishbones; these values of pitch correspond to co-passing FI. In run O70, these ranges were $\SI{60}{keV} \leq E \leq \SI{70}{keV}$ and $0 \leq p \leq 0.7$, corresponding to trapped and co-passing FI. In addition to the fishbone model, a small anomalous diffusivity of \SI{0.15}{m^2.s^{-1}} was applied to FI with $E\leq\SI{45}{keV}$ during the fishbone phase in both simulations; $D_\mathrm{an}$ decreased linearly with energy from \SIrange{0.150}{0.105}{m^2.s^{-1}} between \SI{45}{keV} and \SI{75}{keV}. Grey and red shaded rectangles indicate the periods, before and after the first two fishbones, over which synthetic FI distributions and measured NC and FIDA signals were averaged to produce the subsequent figures in this section.}
\label{fig:FCfish}
\end{figure}

Later in each of the discharges from this set, a series of large fishbones caused transient FI redistribution which the simple diffusive transport model is unable to capture. This redistribution resulted in drops of $\sim10\%$ in the global neutron rate, as seen in figure \ref{fig:FCfish}.
As mentioned earlier, \textsc{Nubeam} contains a fishbone model which allows FI to be expelled from the plasma within a specified part of velocity space, with a chosen periodicity and characteristic loss time. This model was previously used to model changes in global and local neutron emission in another series of MAST discharges \cite{Klimek2015}, but FIDA data were unavailable for those shots. In the series of discharges studied here, the fishbone model was initially used with settings similar to those applied in the previous study, cutting out all FI with $E\geq\SI{50}{keV}$ and with $-0.5\leq p \leq0.5$. It was found however that despite the relatively successful match to the global neutron rate and the changes in NC count rate profiles achieved using this model, the changes in FIDA signal were not reproduced. This may be understood by referring to the geometric weight functions of the FIDA system discussed in section \ref{sec:FIDs}. The toroidal FIDA views are insensitive to ions with such a small ratio of parallel to perpendicular velocity, since the LOS are almost tangential to the magnetic field at the point of beam intersection in the core of the plasma. By contrast, the large fishbones in these shots are observed to cause significant drops in toroidal FIDA signal at all radii from the magnetic axis to the plasma boundary. This observation informed the choice of energy and pitch boundaries in the fishbone model within which the FI were removed. An iterative process was adopted by which the simulation parameters were at first coarsely modified to converge on the global neutron rate, then modified more precisely to converge on the relative changes in NC and FIDA radial profiles. The neutron rate comparisons from the two most successful simulation runs are shown in figure \ref{fig:FCfish}

\begin{figure}[h]
\centering
\includegraphics[width=0.8\textwidth]{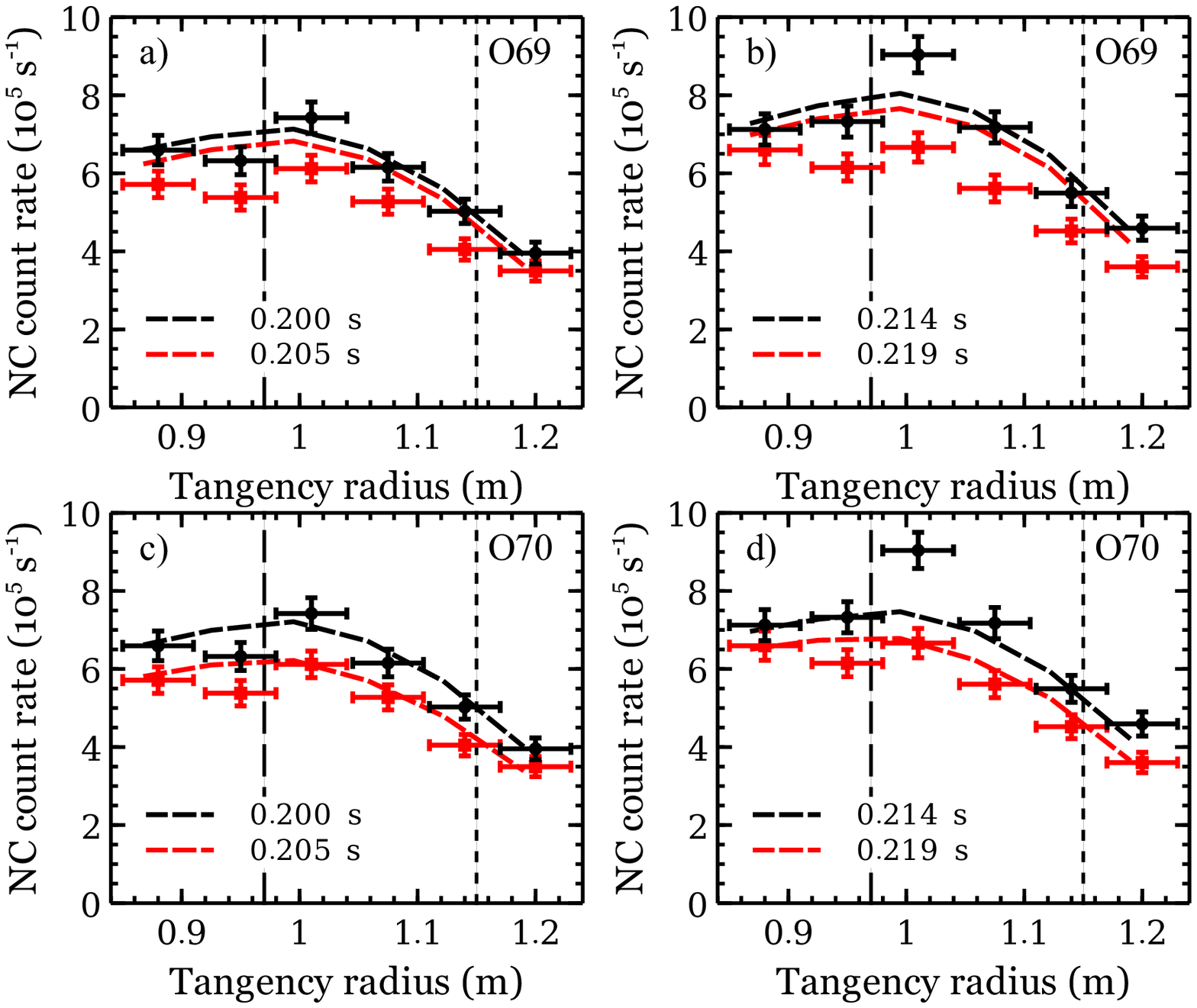}
\caption[Measured and modelled NC profiles affected by fishbones]{Profiles of NC count rate before (black points) and after (red points) the first two fishbones in shots \#29975, \#29976 and \#29980. Panels (a) and (b) compare the measured profiles to those forward-modelled from the neutron emission profiles generated in \textsc{Transp} run O69 for shot \#29976 (dashed lines), while panels (c) and (d) show a comparison with \textsc{Transp} run O70 for the same shot. Vertical dashed and dotted lines mark the radial positions of the magnetic axis and minimum-$q$ surface.}
\label{fig:NCfish}
\end{figure}

Figure \ref{fig:NCfish} shows that \textsc{Transp} run O70, in which high-energy FI with $0 \leq p \leq 0.7$ are removed by the fishbones, produces a good match to the relative size of the observed changes in NC signal across the plasma radius as well as to the shape of the NC profile. Run O69 produces a similar match to the profile shape, but significantly underestimates the size of the drops in signal due to the fishbones.

\begin{figure}[h]
\centering
\includegraphics[width=0.8\textwidth]{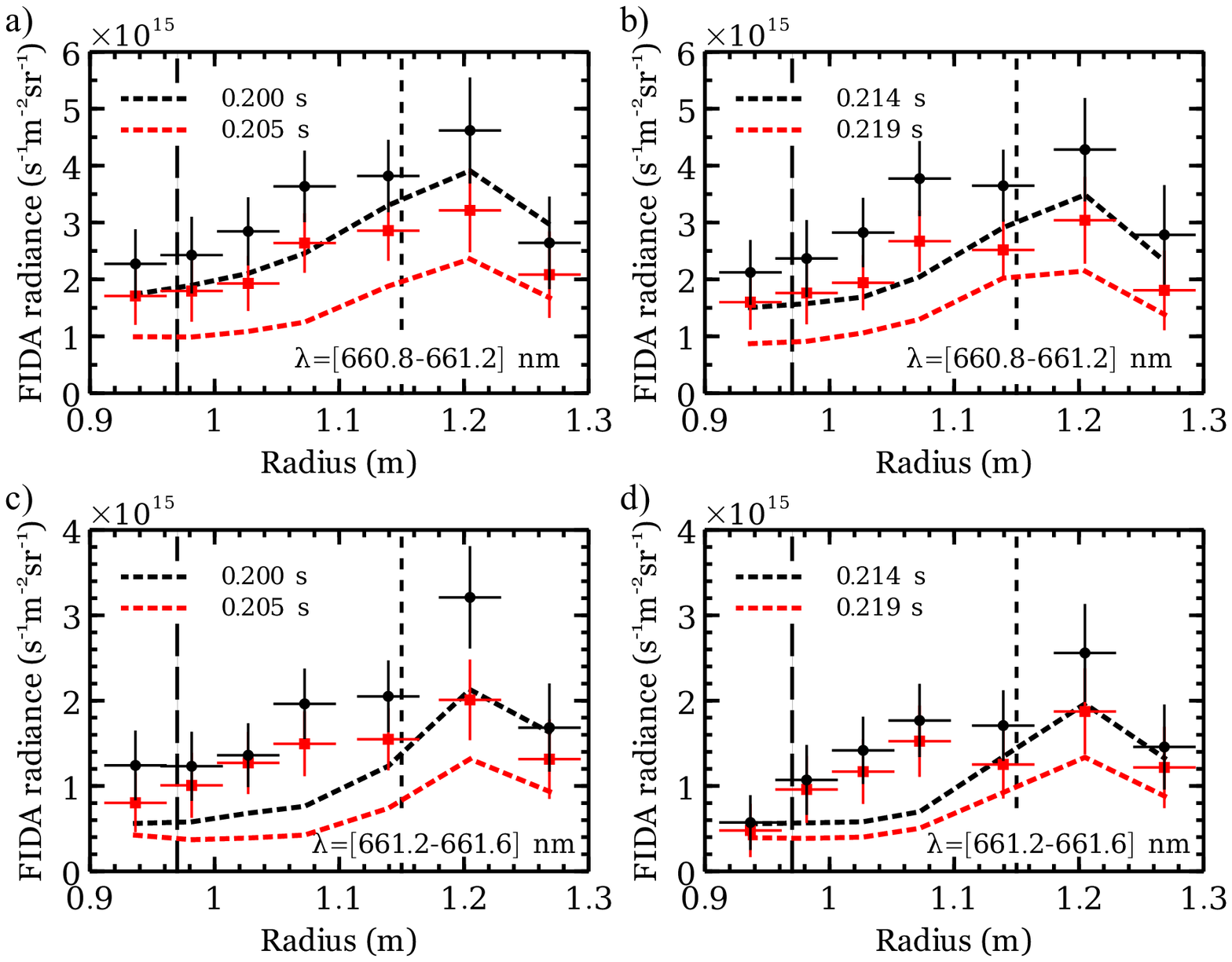}
\caption[Measured and modelled FIDA profiles affected by fishbones]{Profiles of FIDA radiance before (black points) and after (red points) the first two fishbones in shot \#29976. Panels (a) and (b) compare the measured profiles to those forward-modelled with \textsc{FIDAsim} from the FI distributions generated in \textsc{Transp} run O69, in a wavelength range corresponding to a minimum FI energy of \SI{48}{keV} (dashed lines). Panels (c) and (d) show a comparison at higher wavelengths, with $E_\mathrm{min}=\SI{57}{keV}$. Vertical dashed and dotted lines mark the radial positions of the magnetic axis and minimum-$q$ surface.}
\label{fig:FIDAfish}
\end{figure}

\begin{figure}[h]
\centering
\includegraphics[width=0.8\textwidth]{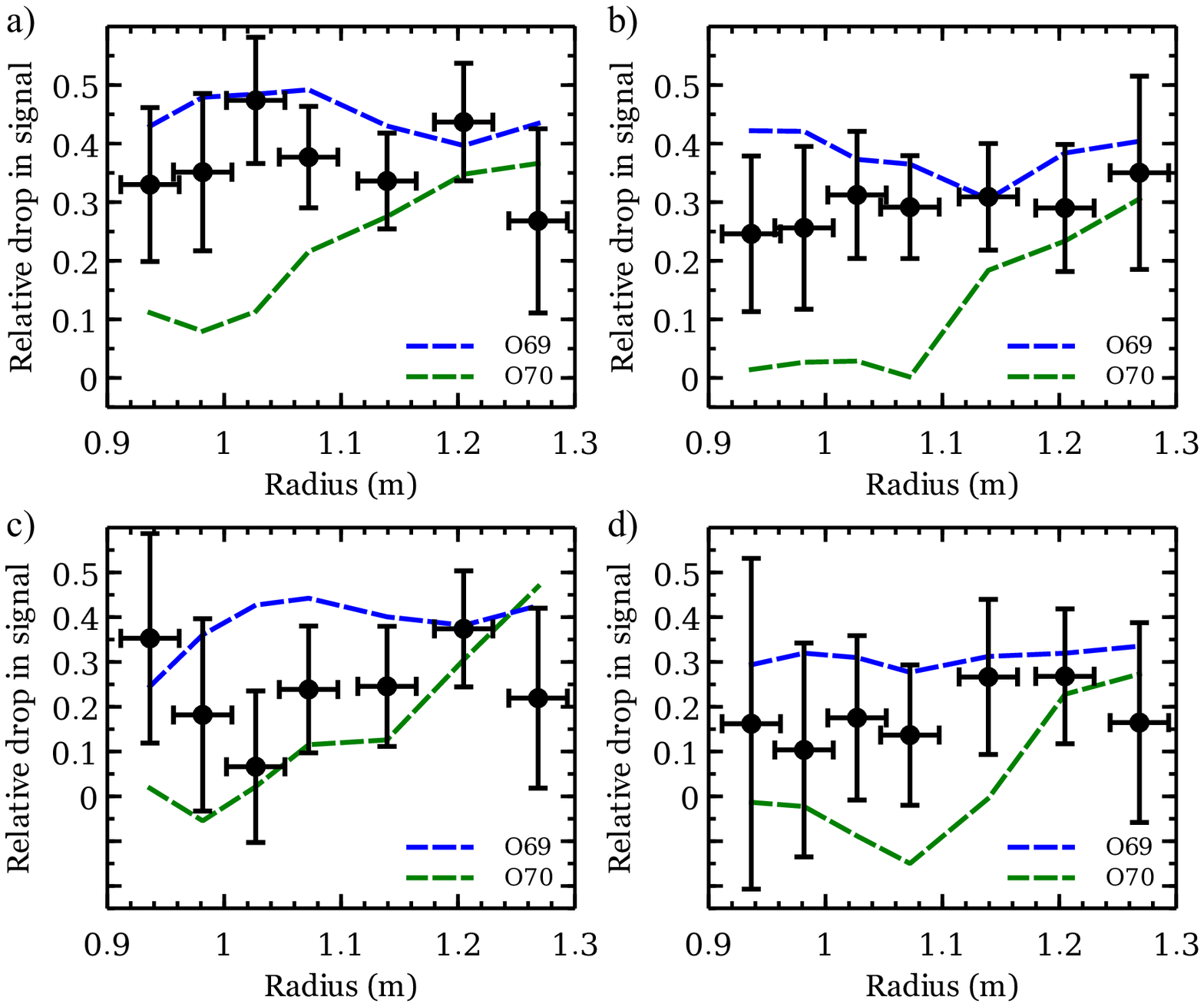}
\caption[Relative changes in FIDA profiles caused by fishbones]{Profiles of the relative reduction in FIDA radiance, defined as $(S_i - S_f)/S_i$, where $S_i$ and $S_f$ are the pre and post-fishbone signals, caused by two fishbones in shot \#29976. The measured relative changes are compared to those derived from the \textsc{FIDAsim} profiles based on \textsc{Transp} run O69 (blue dashed lines) and O70 (green dashed lines). Wavelengths are the same as those in the corresponding panels of figure \ref{fig:FIDAfish}; panels (a) and (c) correspond to the first of the modelled fishbones, while (b) and (d) correspond to the second fishbone.}
\label{fig:FIDArelFish}
\end{figure}

The same pair of \textsc{Transp} runs was used to generate synthetic FIDA profiles. Although the absolute match between the modelled and measured signal levels was slightly better for run O70, the simulated fishbones with these settings failed to cause the observed reduction in core FIDA signal. Figure \ref{fig:FIDAfish} shows that run O69,
by contrast, matches the relative change in signal fairly well and also matches the profile shape well, at least at mid-high energies (top panels), even though it underestimates the magnitude of the signal. The comparison between the two runs in terms of their match to the relative change in FIDA signal is quantified in figure \ref{fig:FIDArelFish}. Clearly the removal of high-energy passing ions reproduces the FIDA observations. The resonant interaction of passing FI with fishbones finds a precedent in results from the PBX tokamak \cite{Heidbrink1986}, and was explained theoretically by Betti and Freidberg \cite{Betti1993}.

The fact that run O69 systematically slightly overestimates the magnitude of the changes in FIDA signal across the plasma radius, while run O70 underestimates these changes in the core but matches them well at the edge, suggests that the real effect of the fishbones is best represented by a combination of these models. It is reasonable to suppose that the properties which allow FI to resonate with the mode, and hence to be strongly redistributed, are a function of radial position in the plasma. This dependence cannot be captured with the simple \emph{ad hoc} model available in \textsc{Nubeam}. More realistic simulations must await the inclusion of a first-principles model of resonant transport in the global transport simulations.

\begin{figure}[h]
\centering
\includegraphics[width=\textwidth]{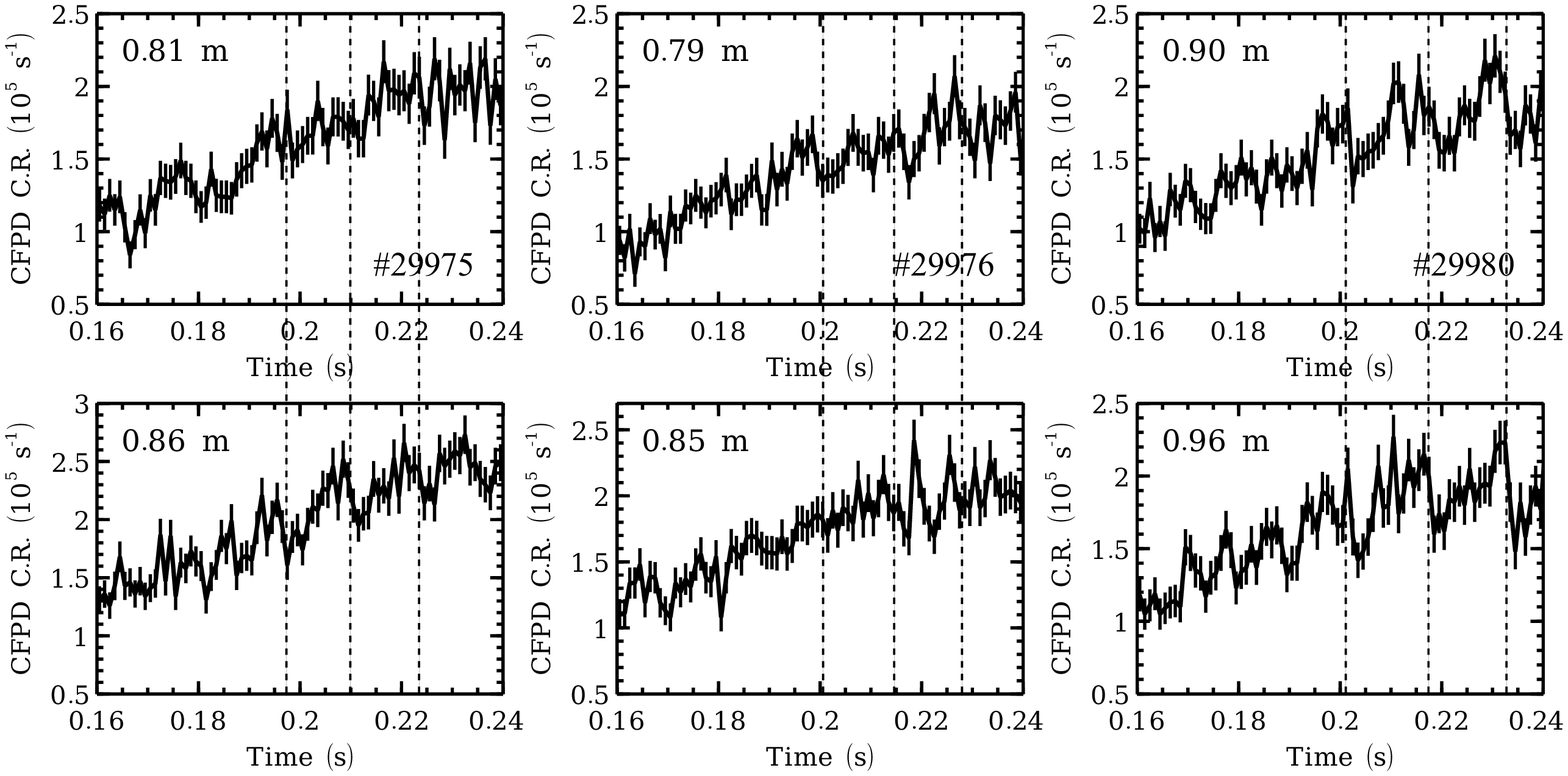}
\caption[CFPD time traces during large fishbones]{Time traces of data from two channels of the CFPD during the large fishbones in shot \#29975 \emph{et seq.}. Each column includes data from a single shot, with dotted vertical lines marking the time at which the amplitude of each of the three largest fishbones reached a maximum. The label in each panel indicates the midplane intersection radius of the centre of the bundle of fusion proton trajectories which reaches the detector. The magnetic axis is located at approximately $R=\SI{0.97}{m}$.}
\label{fig:CFPDscan}
\end{figure}

In the set of shots commencing with \#29975, the radial position of the CFPD was scanned between discharges. This allowed the effects of the large fishbones on the CFPD count rate to be determined as a function of midplane intersection radius; the data from each channel are shown in figure \ref{fig:CFPDscan}. It is apparent that the strongest effect of the fishbones on the count rate is observed on trajectories intersecting the midplane close to the magnetic axis. Channels with midplane intersection radii from \SIrange{0.86}{0.96}{m} see a strong depletion in the count rate coinciding with each of the large fishbones, while those at \SI{0.79}{m} and \SI{0.81}{m} see only a weak effect. Without a reliable synthetic diagnostic, the fishbone modelling carried out with \textsc{Transp/Nubeam} cannot be validated against these measurements. Nonetheless, the observation of fishbone-induced drops in signal in \emph{all} of the traces in figure \ref{fig:CFPDscan} is broadly consistent with the NC and FIDA measurements, which show a significant reduction in signal across the core of the plasma.

\section{Summary and conclusions}
\label{sec:Summary}
The results presented in this paper demonstrate the advantages of employing several complementary diagnostics to study the behaviour of the FI population. Extending previous studies on MAST \cite{Cecconello2015, Jones2013, Turnyanskiy2013, Keeling2015}, a systematic analysis of NC and FIDA data has allowed the deleterious effects of chirping TAE and fishbones on the confined FI population to be identified with confidence. The velocity-space sensitivity of the FIDA measurements moreover permits losses of FI to be identified unambiguously, and these losses are seen to be strongly correlated with the MHD activity.

Radial profiles of neutron and FIDA emission have been used to constrain transport modelling and to establish the effects of fishbones on the FI distribution. A simultaneous comparison of the modelling results to measurements made with the FC, NC and FIDA diagnostic has allowed constraints to be placed on the effects of fishbones on the velocity-space distribution of FI; these modes are seen to affect high-energy, passing particles in the core of the plasma. Data from the prototype CFPD support the conclusion that fishbones have a strong impact on core FI confinement.

Resonant energetic-particle-driven MHD instabilities observed in MAST, as discussed in section \ref{sec:MHD}, include chirping TAE and fishbones. In this work, observations of changes in the FI diagnostic signals correlated with each of these instabilities were presented and the interpretation of these observations was discussed. The results of \textsc{Transp} modelling in which anomalous FI transport or loss processes were invoked to try to match the observations were presented and compared with the measurements. Four main conclusions are supported by the results presented here:
\begin{itemize}
\item{Drops in both FIDA and NC signals are correlated with chirping TAE and fishbones at a statistically significant level, indicating a reduction of the confined FI density.}
\item{Chirping TAE and fishbones cause enhanced losses of FI from the plasma.}
\item{The profiles of NC and FIDA signals averaged over individual bursts of MHD activity are well modelled by applying anomalous diffusion to the FI in \textsc{Nubeam}.}
\item{Fishbones strongly affect the high-energy, passing FI population in a manner which may be reproduced with some success using the fishbone model in \textsc{Nubeam}.}
\end{itemize}

The additional flexibility offered by the NBI system on the forthcoming MAST-Upgrade device, which will have the capability to inject beams simultaneously on and off-axis, will allow a detailed investigation into the relationship between beam deposition profiles, MHD mode activity and FI transport \cite{Keeling2015}.

\ack This work was part-funded by the RCUK Energy Programme under grant EP/I501045, by the Swedish Research Council, and by EURATOM. To obtain further information on the data and models underlying this paper, please contact PublicationsManager@ccfe.ac.uk. The views and opinions expressed herein do not necessarily reflect those of the European Commission.

\section*{References}
\bibliographystyle{unsrtmod}
\bibliography{library}

\end{document}